\documentclass[useAMS,usenatbib]{mn2e}

\usepackage[dvips]{graphicx}
\usepackage{verbatim}
\usepackage{lscape}
\usepackage[dvips]{color}
\usepackage{wasysym}
\usepackage{float}

\title[High Frequency VLBI Rotation Measure of 8 AGN]{High Frequency VLBI Rotation Measure of 8 AGN}

\author[J. C. Algaba]{J. C. Algaba$^{1}$\thanks{E-mail: algaba@asiaa.sinica.edu.tw} \\
$^{1}$Academia Sinica, Institute of Astronomy and Astrophysics, P.O. Box 23-141, Taipei 10617, Taiwan, R.O.C. \\}
\begin{document}

\date{Accepted. Received; in original form}

\pagerange{\pageref{firstpage}--\pageref{lastpage}} \pubyear{2011}

\maketitle

\label{firstpage}

%

\begin{abstract}
We have studied Very Long Baseline Array (VLBA) polarimetric observations of 8 sources including quasars and BL Lacs at 12, 15, 22, 24 and 43 GHz and high frequency rotation measure ($RM$) maps are presented. We find typical values for the $RM$ in the VLBI core of several thousand rad/m$^2$, which are higher than values in the literature at lower frequencies. Assuming a dependence of the form $RM\propto \nu^a$, we obtain an average value of $a=3.6\pm1.3$, which is larger than that expected by theoretical considerations. Rotation measures are detected in the jet of only two sources and we find that only 0906+430 (and possibly 1633+382) show indications of a robust gradient. We discuss the Faraday--corrected polarization properties of the sources. Our interpretation supports the presence of helical magnetic fields with new, unresolved, components affecting the intrinsic direction of polarization close to the base of the jet of some objects.

\end{abstract}


\section{Introduction}
The emission associated with jets in active galactic nuclei (AGN) is due to synchrotron radiation over a wide range of the electromagnetic spectrum, including radio frequencies. The emission is generally highly linearly polarized, more than 50\% in some optically thin jet regions \citep[see e.g.][]{ListerHoman}, clearly indicating the presence of highly ordered magnetic fields. In several cases it has been found that this polarization is perpendicular to the direction of the jet \citep[][and references therein]{Gabuzda2000} or has a spine/sheath structure \citep{Attridge99, Pushkarev05}; see also numerous examples in the MOJAVE database \citep{ListerHoman}.

Although this can be interpreted as a series of shocks in the jet enhancing the perpendicular component of the magnetic field, and/or the interaction of the jet with the surrounding media \citep{Laing08}, the presence of polarization either parallel or perpendicular to the jet can also be easily explained with a helical magnetic field. 
There is evidence that helical magnetic fields seem to be present in, at least, a fraction of AGN. \cite{Gabuzda2006} summarizes different observational polarization tests in order to distinguish between helical magnetic fields and alternative explanations concluding that there are a number of sources for which reasonable evidence for helical jet magnetic fields is available.

One way to study the structure of the intrinsic magnetic field is the analysis of the rotation measure ($RM$) distribution across the source. The polarization angle $\chi$ rotates following the relation $\chi=\chi_{0}+RM\lambda^2$, where $\chi_{0}$ is the intrinsic polarization angle, $\lambda$ is the wavelength and $RM$ is the rotation measure, given by
\begin{equation}
RM\propto\int n_e B \mbox{\textperiodcentered} dl, 
\end{equation}
where $n_e$ is the electron density and $B\mbox{\textperiodcentered} dl$ is the magnetic field along the line of sight. 

The first implication is that the intrinsic polarization (and hence the inferred direction of the magnetic field giving rise to it) will be altered by the effects of the Faraday rotation. As the $RM$ depends on the component parallel to the line of sight of the magnetic field, we will have, in general, a variable alteration of the polarization direction across the source. Thus, if we want to properly understand the intrinsic properties of the magnetic field, we need to map the $RM$ across the source in order to adequately subtract its effects in the different regions.

If we have a helical magnetic field, its toroidal component will produce a rotation measure change across the jet, giving rise to a gradient \citep{Blandford93}. In the simplest case, with the jet perpendicular to the line of sight in the observer's rest frame, we would observe an antisymmetric distribution of the rotation measure, with positive values on one side of the jet, negative values on the other and a null $RM$ in the centre. As the viewing angle decreases, there will be an offset of the absolute value of the $RM$ values and, if the viewing angle turns out to be smaller than the pitch angle, we will observe only positive values for the rotation measure \citep{Uchida04,Asada02}. 

Such $RM$s have already been detected: \cite{ZT03, ZT04} studied $RM$s in a sample of AGNs. \cite{Asada02,Asada08} found a time-variable $RM$ gradient over more than 100pc along the jet in 3C273, the origin being quite probably the sheath around the ultra-relativistic jet. \cite{Gabuzda04} also found rotation measure gradients ranging from negative to positive values across the jet in 0745+241. However, the stability of the $RM$ structures is still unclear: \cite{ZT01} found time variability of the $RM$ in 3C273 and 3C279 but \cite{Gomez11} found that, although there were changes in the linear polarization of 3C120, the underlying rotation measure remained unaltered along 2 to 5 mas from the core over several years.

Previous results are based on observations obtained with the Very Long Baseline Array (VLBA) using, with the exception of few cases, low frequency bands (typically in the range of 8--15 GHz) based on a compromise between resolution and sensitivity, and only \cite{Gomez11} and \cite{Attridge05} have obtained results for frequencies as high as 86 GHz for the radio galaxy 3C120 and 3C273, respectively. Thus, we seek to obtain more measurements at higher frequencies so that we can study the behavior and structure of the rotation measure closer to the base of the jet in a larger number of cases. We can then compare our results with the ones obtained at lower frequencies.

In this paper we study the rotation measure of 8 sources observed with the VLBA from 12 to 43 GHz. In section 2 we summarize the observations and data reduction. In section 3 we present our results. In section 4 we discuss our findings and we present our conclusions in section 5. An analysis of the robustness of rotation measure gradients is presented in the appendix.

\section{Observations and Data Reduction}
Polarization observations of 6 radio--quasars (0133+476, 0420-014, 0745+241, 0906+430, 1633+382 and 1954+513) and 2 BL Lac objects (0256+075 and 1823+568) were carried out in a 24 hour session on November 2, 2008 using the VLBA. The frequencies selected were  12.039, 15.383, 21.775, 23.998 and 43.135 GHz, each with 2 IFs, with a bandwidth of 8MHz and a bitrate of 128MB/s. The sources were observed in a ``snapshot'' mode, with 8--10 several minute scans for each frequency and object spread in time, so that the resulting UV coverage was quite uniform. The data reduction and imaging was done with the NRAO Astronomical Image Processing System (AIPS) using standard techniques. The reference antenna used was Los Alamos. Simultaneous solutions for the instrumental polarizations and source polarizations for the compact source 1954+513 were derived using the AIPS task LPCAL. 

{We calibrated the electric vector position angles (EVPAs) using the VLBA D-Terms (G\'omez et. al, 1992). Comparison of the D--terms against a set of tabulated values (previously calibrated by other means) proves to be a reliable method for calibrating the absolute L--R phase offset in VLBA observations. We were able to obtain two sets of calibrated data with either the same or nearly the same frequencies (S. P. O'Sullivan, private communication; A. Reichstein, private communication) so that it was possible for us to apply this method to our data.}

{To ensure the reliability of this method, we performed a series of extra checks. First, we compared the independent results given by the two comparison sets of D--terms for consistency. Second, we compared our new images with our previous images (Algaba 2010). Third, for some frequencies, we were also able to compare the resulting polarization angles with images from the MOJAVE\footnote{http://www.physics.purdue.edu/MOJAVE/}, University of Michigan Radio Astronomy Observatory Database\footnote{http://www.astro.lsa.umich.edu/obs/radiotel/umrao.php and M. Aller, personal communication} and the 7-mm monitoring by Boston University Blazar Group \footnote{http://www.bu.edu/blazars/VLBAproject.html} databases. Fourth, we obtained the EVPAs of several sources in the optically thin jet regions and checked for consistency with modest Faraday rotation. We estimate the overall error in the EVPA calibration to be about $4^{\circ}$ for all frequencies.}

After the initial construction of total intensity maps, we made new versions at 12, 15, 22, 24 and 43 GHz using the best calibrated data but convolving all frequencies with a circular version of the beam obtained for the 15 GHz map. This beam was chosen as a compromise between the lower (12 GHz) and higher (43 GHz) resolution maps, super--resolving 12 GHz only by a modest amount. We then made the maps for the Stokes parameters $Q$ and $U$ that we used to construct the polarized flux ($p=\sqrt{Q^2+U^2}$) and the polarization angle ($\chi=\frac{1}{2}\arctan\frac{U}{Q}$), with their respective noise maps, using the AIPS task COMB. {As information about the absolute position is lost during the calibration process, and given the core shift at different frequencies \citep{Lobanov}, we had first to properly align the different maps at various bands. For this, we used the program developed by \cite{Croke08} based on the cross--correlation of optically thin regions of the jet.}

For the construction of the rotation measure maps, as the current version of the AIPS task $RM$ is limited by a maximum of 4 input frequencies, we used a modified version obtained from R. Zavala that can construct rotation measure maps using up to 10 frequencies. The distribution of $RM$ was obtained in the regions where the error in the input maps was smaller than 15\degr. We found that the results are very similar compared to when we blank those regions with the error on the $RM$ exceeding 100 rad/m$^2$, except the blanking edges appears smoother in the former case. In order to check the goodness of the $\lambda^2-$fit we used the AIPS task RMCUB, also included in R. Zavala's package, which allows us to plot a grid of $RM$ fits and obtain fit parameters, such as $RM$ fit error, $\chi^2$, correlation coefficient and the Q--value.

For 0420--014, 0745+241, 1823+568 and 1954+513 we used all 5 available frequencies. The case of 0133+476 is very complex \citep[see][]{Algaba12}, including possible integration of different polarization components or frequency variability of the rotation measure, and thus we decided to only fit our three highest frequencies. In the cases of 0256+075 and 0906+430, the polarization structure becomes more complicated when reaching the highest frequency in our observation. Consequently, the 43 GHz EVPA value significantly departs from the trend seen in other frequencies. This indicates that we are either probing a different region or that we are integrating over different non--resolved components. Hence, we consider these values not to be representative of the region we study in the rest of the frequencies and so we decided to not include 43 GHz in our $RM$ maps. For 1633+382 the 43 GHz angles have larger errors, thus we exclude this frequency from our study. Frequencies used in each source for the calculation of $RM$ maps are summarized in Table 1.

There are a number of factors which might alter the orientation of the polarization angles in our maps. For example, a transition in the optical depth of the source can induce a 90\degr change in the polarization angle. Also, 180\degr ambiguities are inherent to the polarization angles. We investigated these in \cite{Algaba12}, where we analyzed the total and polarized intensity properties of the sources, finding that all sources were either only optically thin or only optically thick in the regions where we find RM. Hence, we are confident that no 90\degr jump due to a thin--thick transition is invalidating our results. In the same way, we studied the possibility of 180\degr ambiguities in the polarization angles in the region close to the core and these have been taken into account in the calculations.

In order to double check our results, various $RM$ maps were created for each source using different blanking methods and/or a different set of frequencies, comparing the fits both by eye and checking the goodness of the fit using the task RMCUB. We noticed that, for sources like 0256+075 this implied the addition of noisy areas in the $RM$ map and, in some cases as in 0133+476, the emergence of patchy areas. 

In cases such as 0256+075 and 0906+430, including 43 GHz resulted the fit not to pass through all the data and tilted the slope to a wrong value. As we excluded 43 GHz in these, the goodness of the fits improved from $R^2\sim$0.35 and 0.86 to $R^2\sim$0.95 and 0.96 for 0256+075 and 0906+430 respectively.
For other sources, such as 1633+382, the $RM$ map did not change significantly, but due to the error associated with the additional frequency, the fit turned out to be worse.

We also checked the consistency of the $RM$ maps when a shift on any of the input maps was applied. For this, we considered input maps with and without a core shift \citep{Lobanov}, as well as horizontal or vertical shifts up to 200$\mu$as, and we found no significant difference in the output $RM$ maps.

\section{Results}

Rotation measure maps derived for all 8 sources studied here are shown in the top panels of Figure 1. Here, the contours correspond to the total intensity of the convolved version of the map for 15 GHz, and the color levels indicate the rotation measure found. Peak and bottom intensity contours, corresponding to $3\times$RMS, are given in Table 1, together with the frequencies used for the Faraday rotation $\lambda^2-$fit, and the rotation measure scale for all sources. In all maps, contours increase in steps of $\sqrt6$, the color code for the rotation measure is shown in the top, and the beam in the bottom left corner of the image.

\begin{figure*}
  \centering
  \begin{tabular}{ccc}
\includegraphics[width=5.5cm, trim=0cm 0.6cm 0cm 0cm, clip=true]{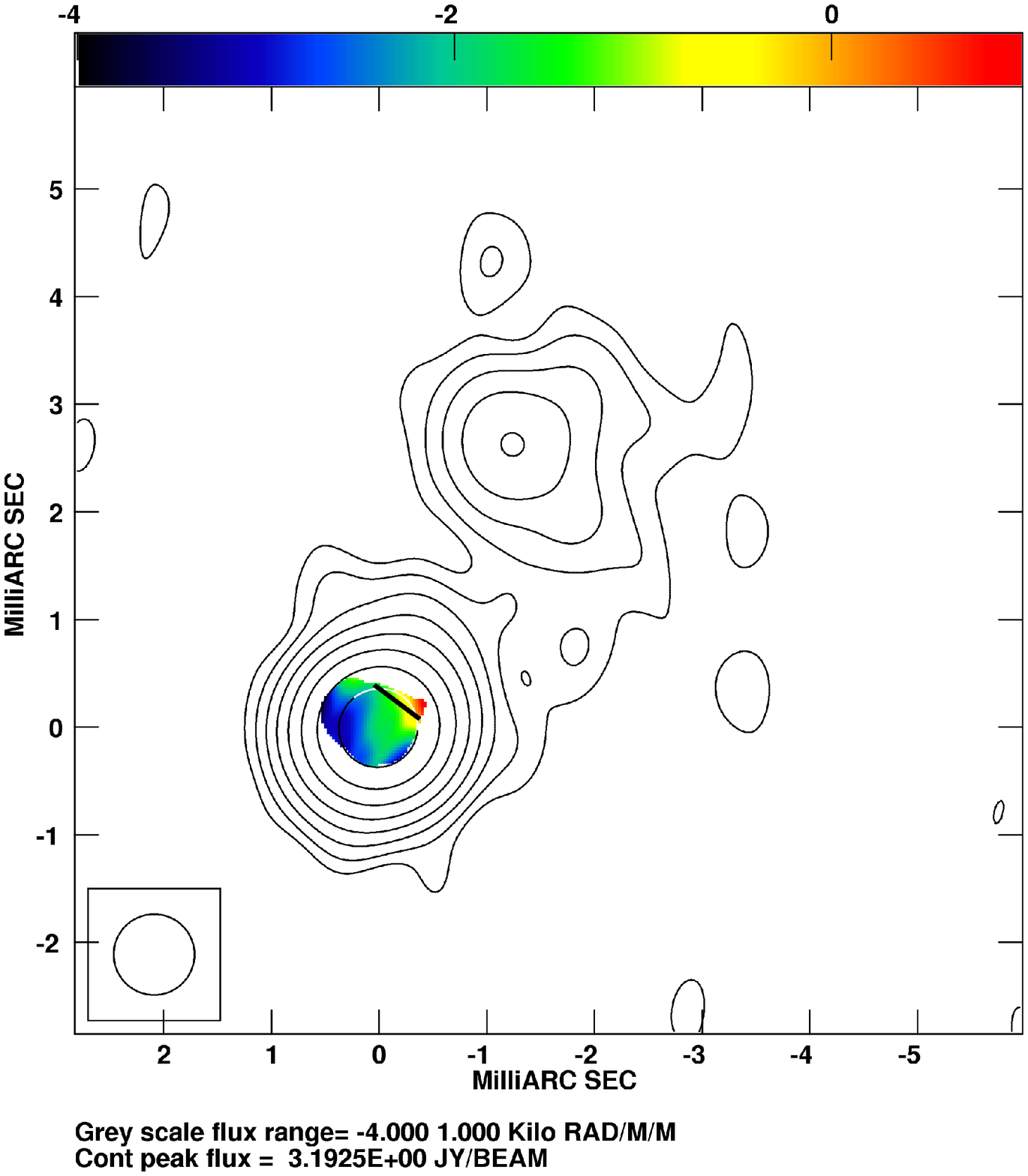}&
\includegraphics[width=5.5cm, trim=0cm 0.6cm 0cm 0cm, clip=true]{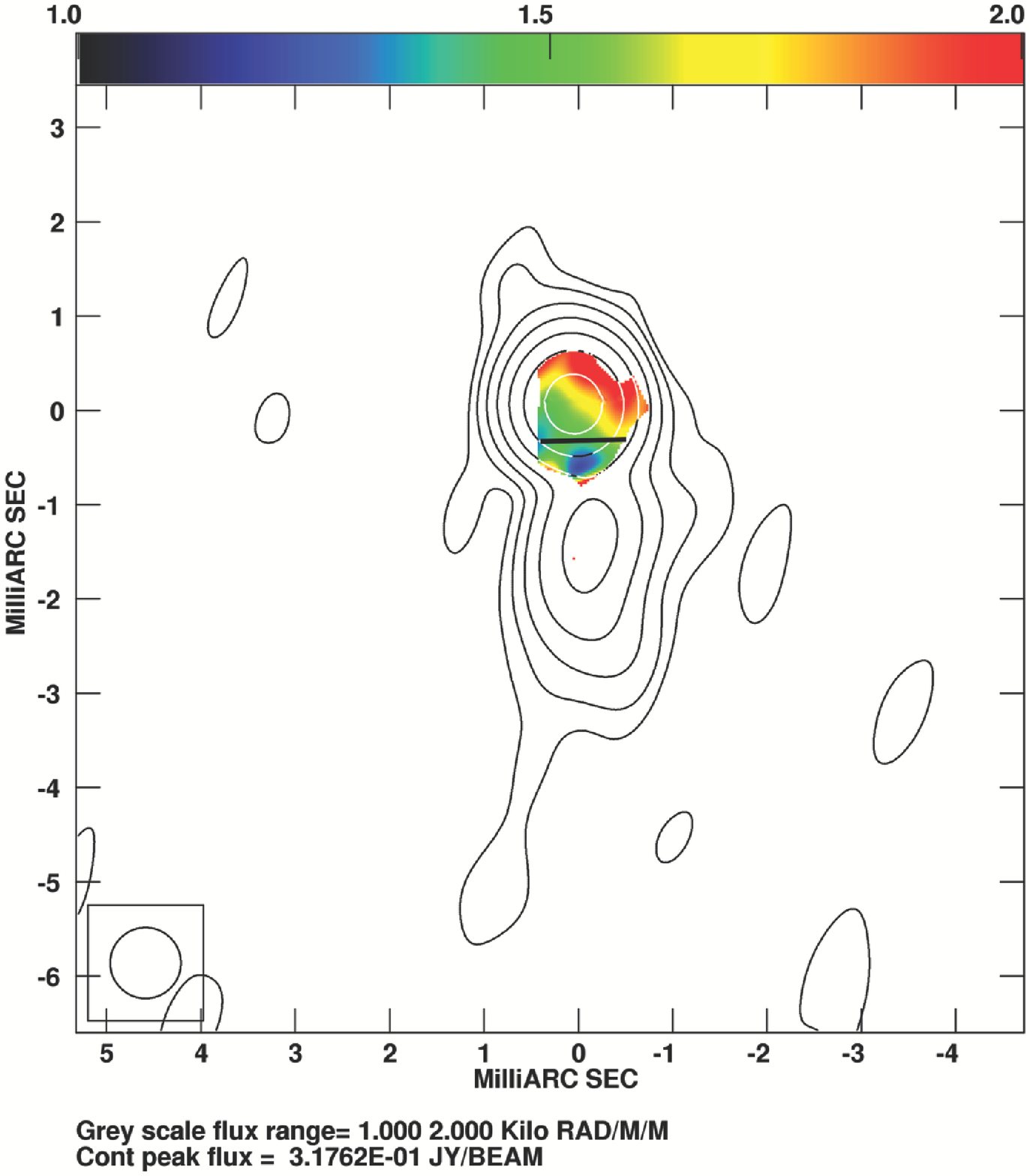}&
\includegraphics[width=5.5cm, trim=0cm 0.6cm 0cm 0cm, clip=true]{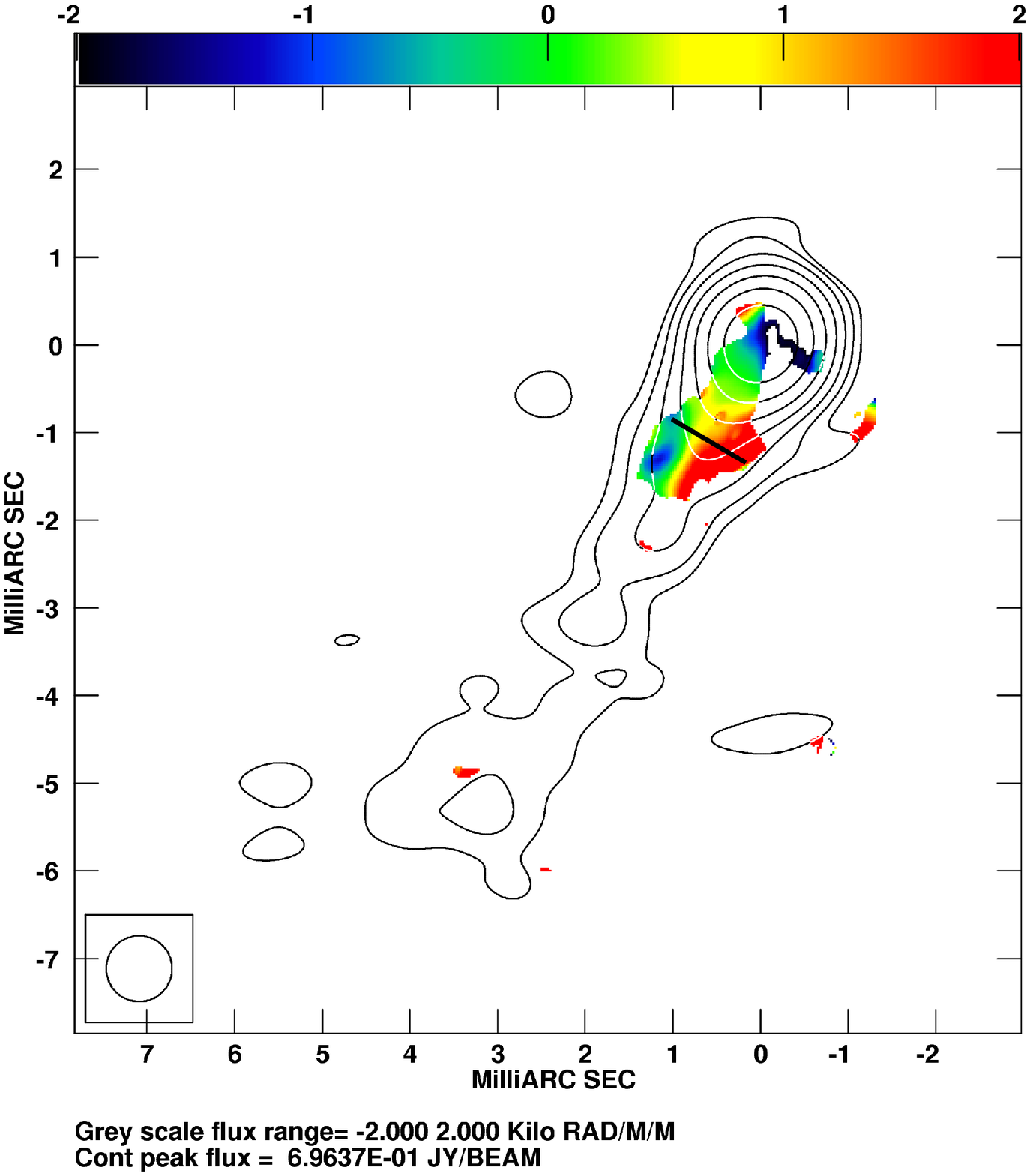}\\

\includegraphics[width=5.5cm, trim=0cm 0cm 0cm 0.9cm, clip=true]{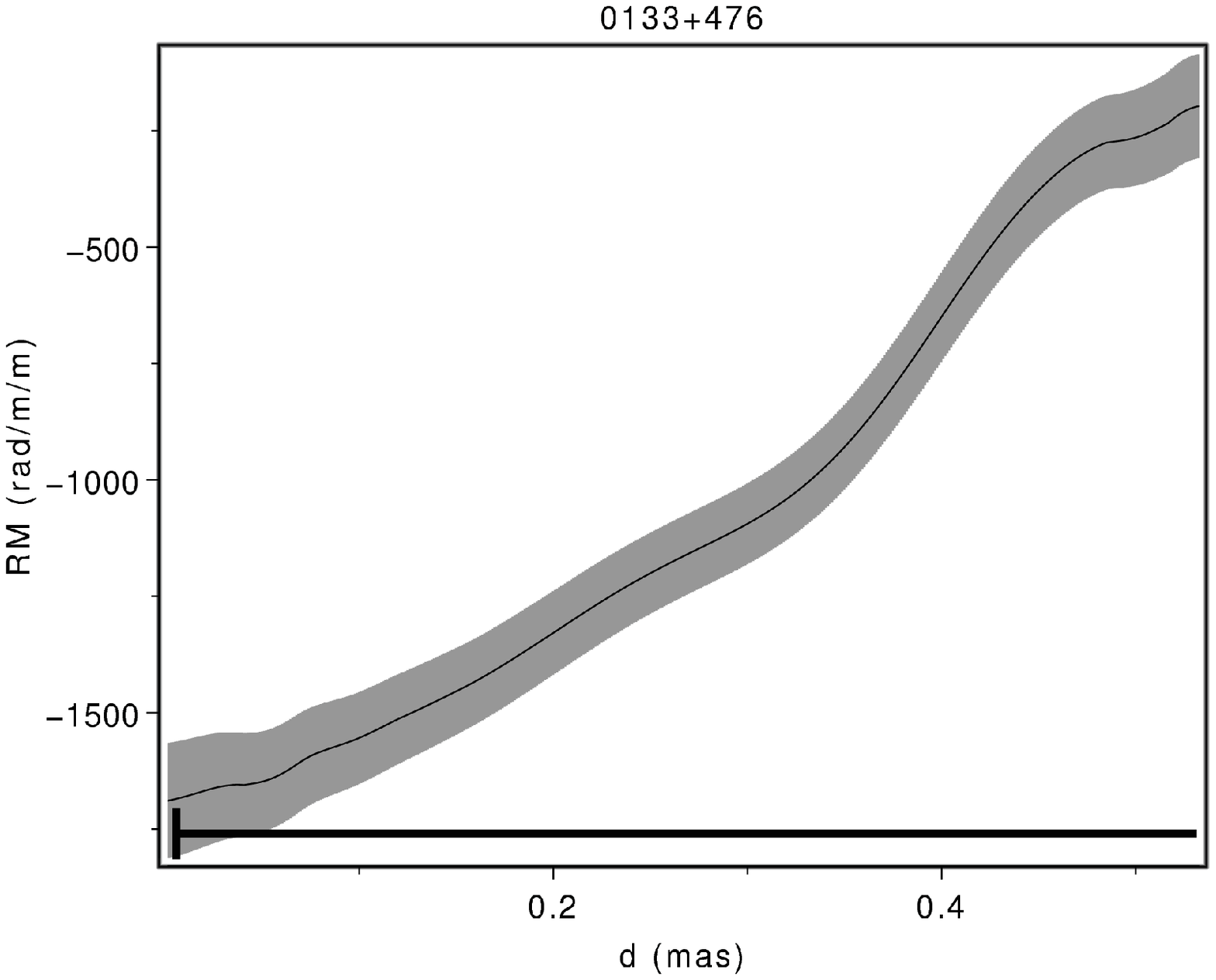}&
\includegraphics[width=5.5cm, trim=0cm 0cm 0cm 0.9cm, clip=true]{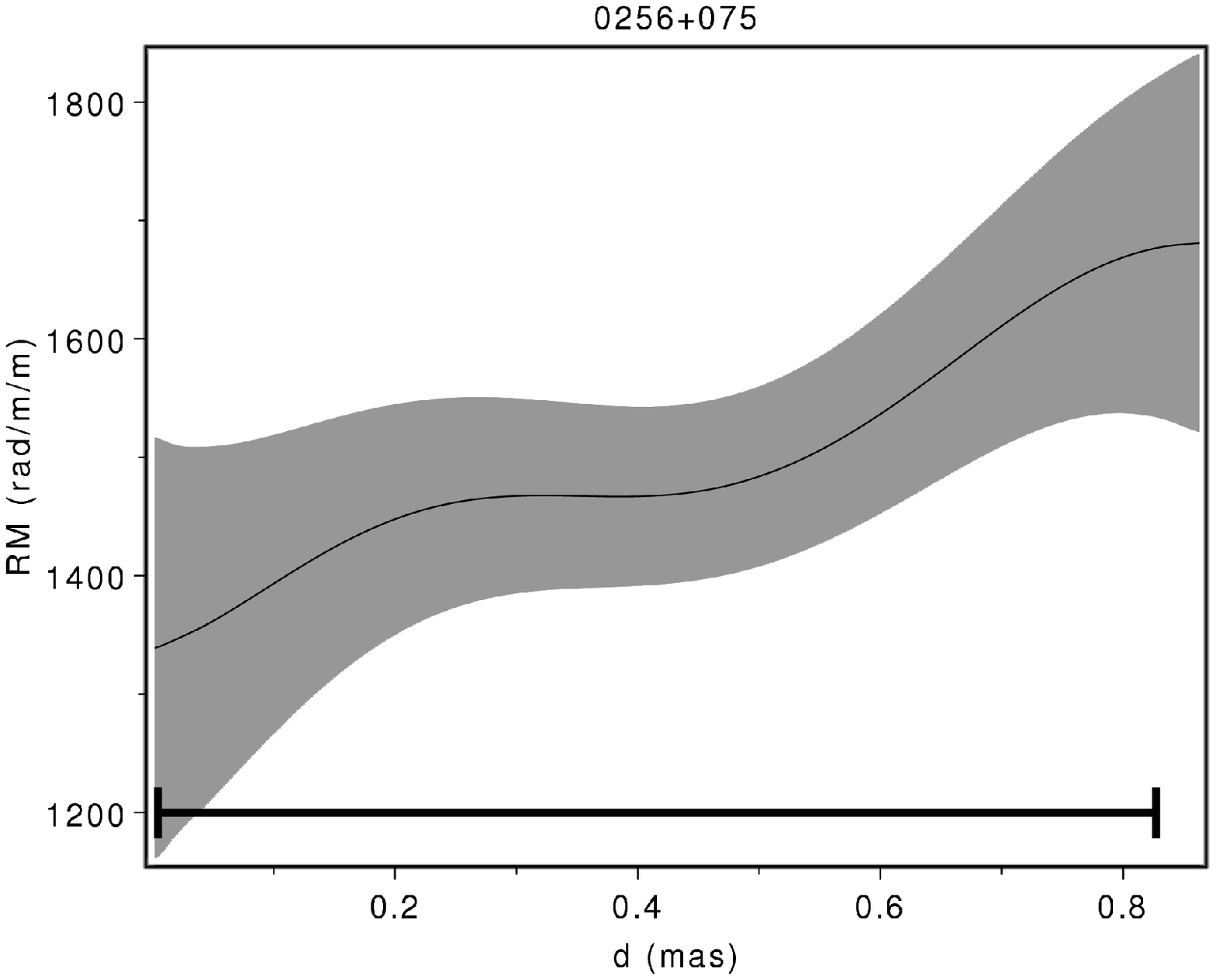}&
\includegraphics[width=5.5cm, trim=0cm 0cm 0cm 0.9cm, clip=true]{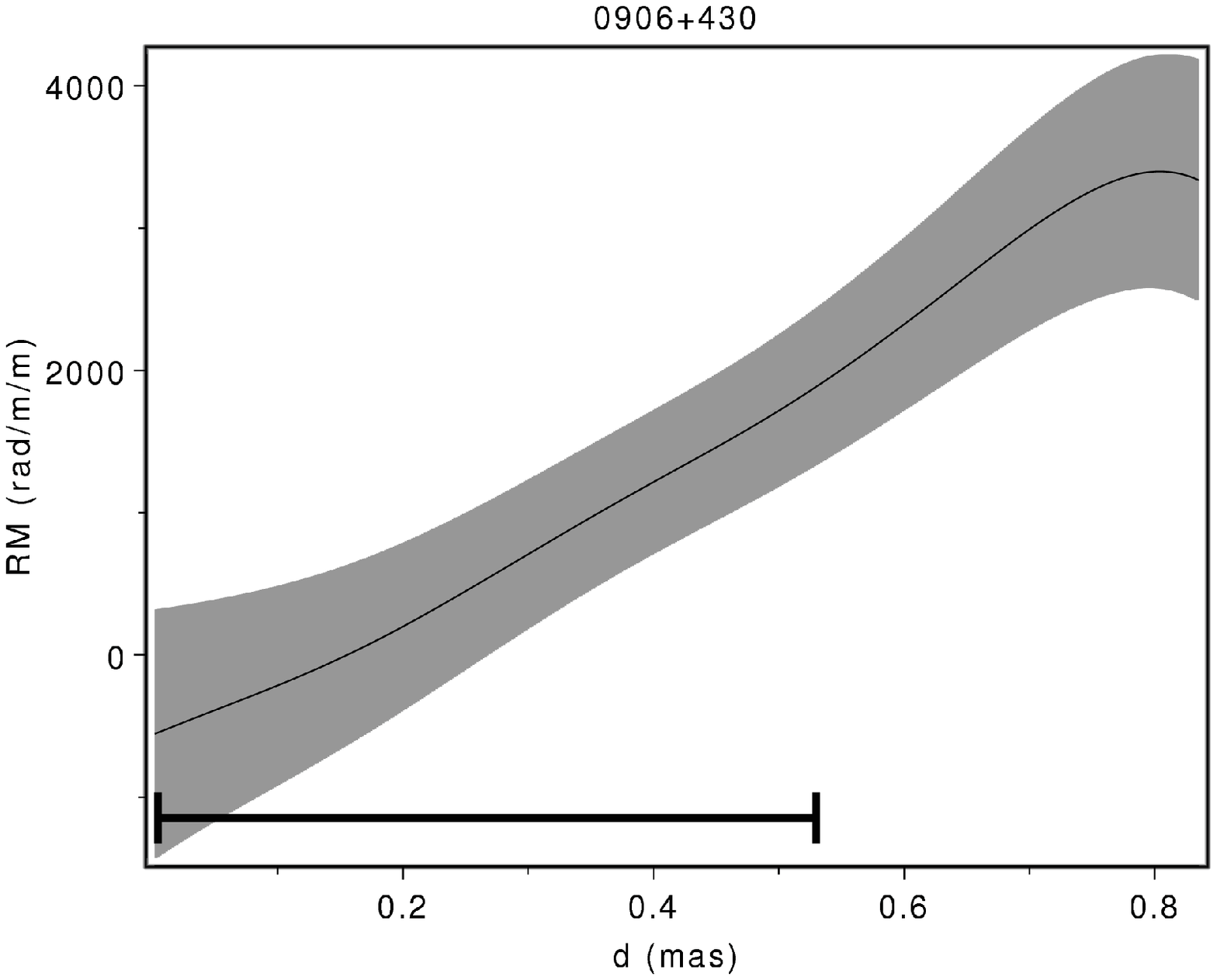}\\

(a) 0133+476&(b) 0256+075&(c) 0906+430\\

\includegraphics[width=5.5cm, trim=0cm 0.6cm 0cm 0cm, clip=true]{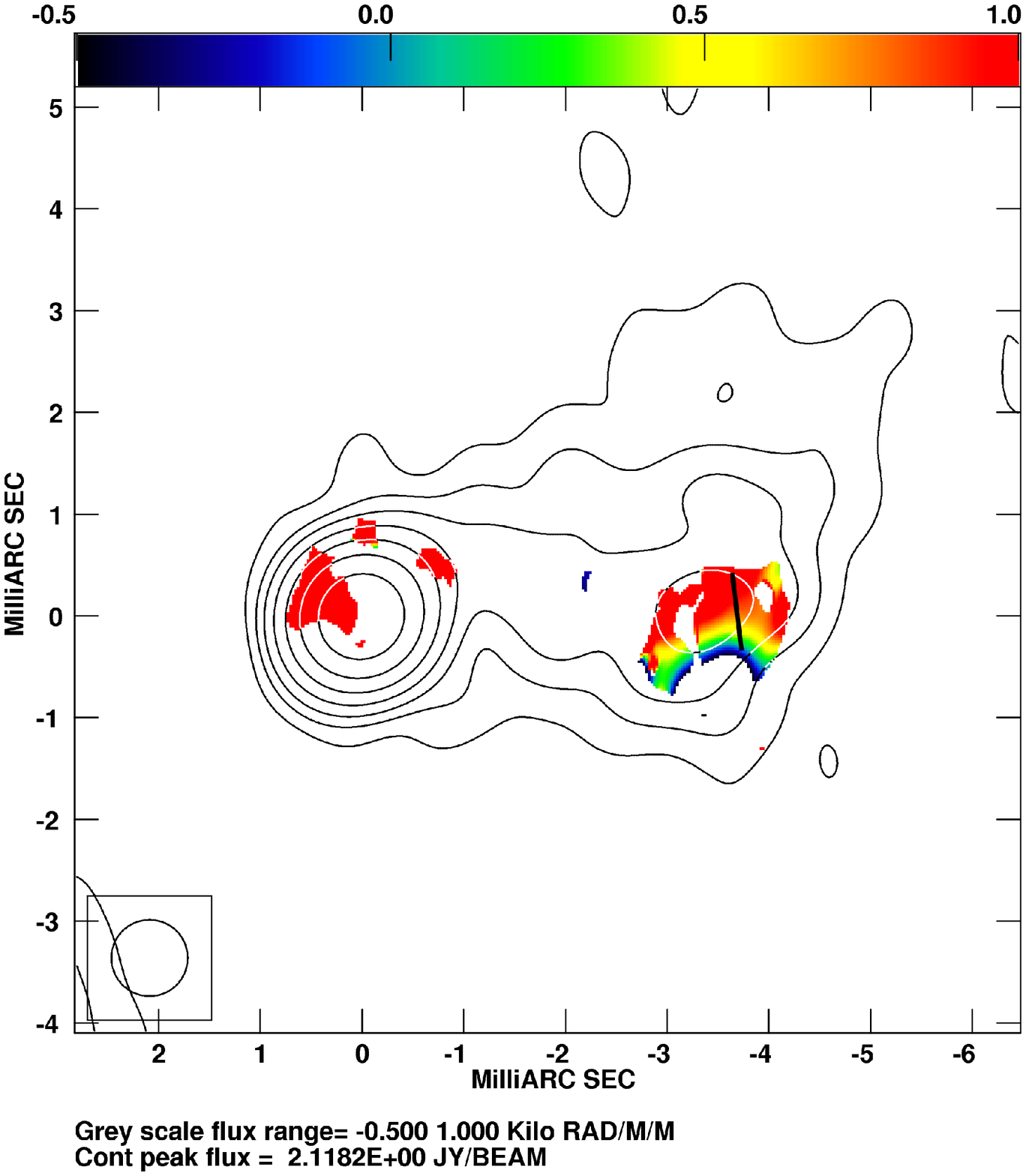}&
\includegraphics[width=5.5cm, trim=0cm 0.6cm 0cm 0cm, clip=true]{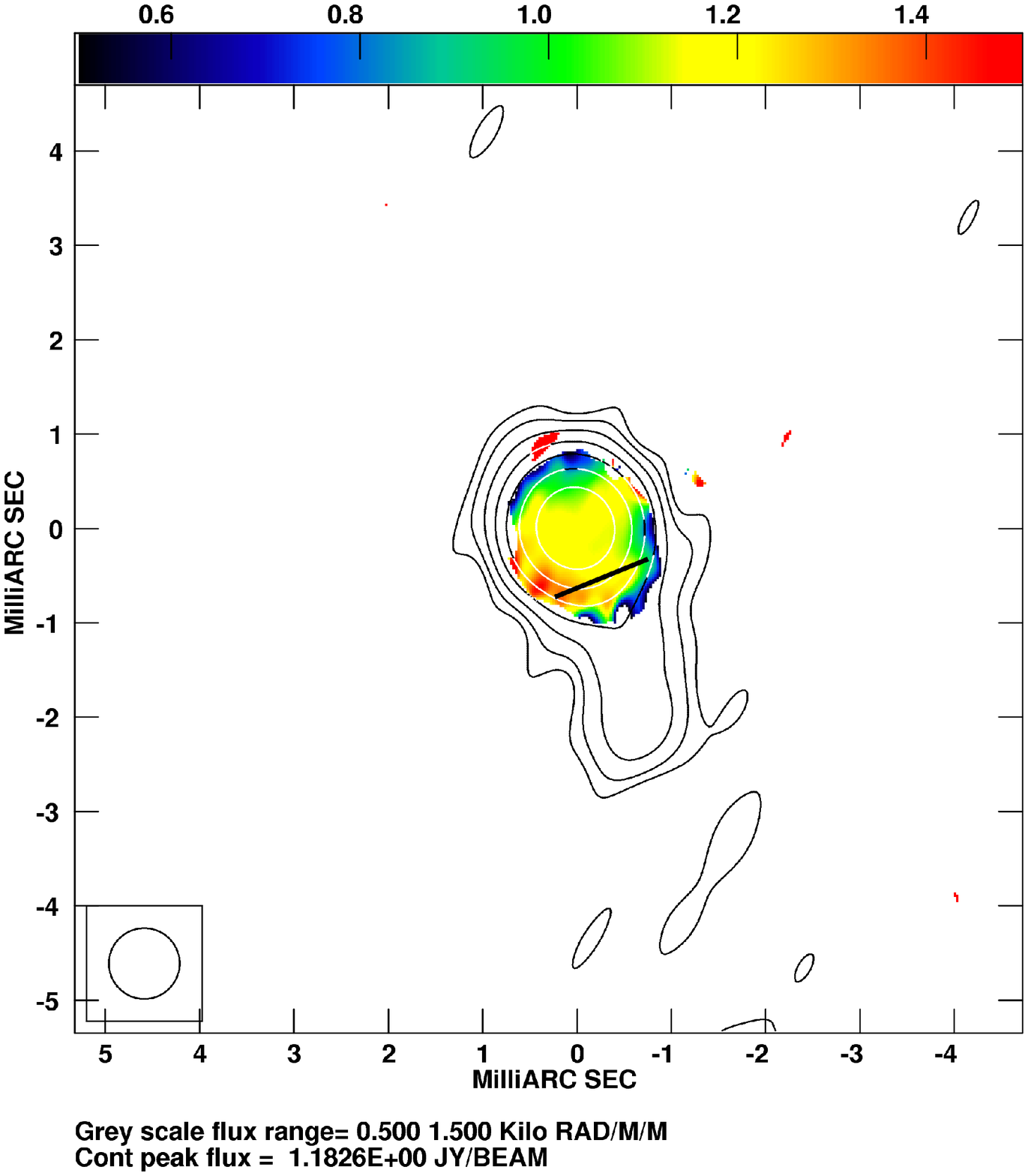}&
\includegraphics[width=5.5cm, trim=0cm 0.6cm 0cm 0cm, clip=true]{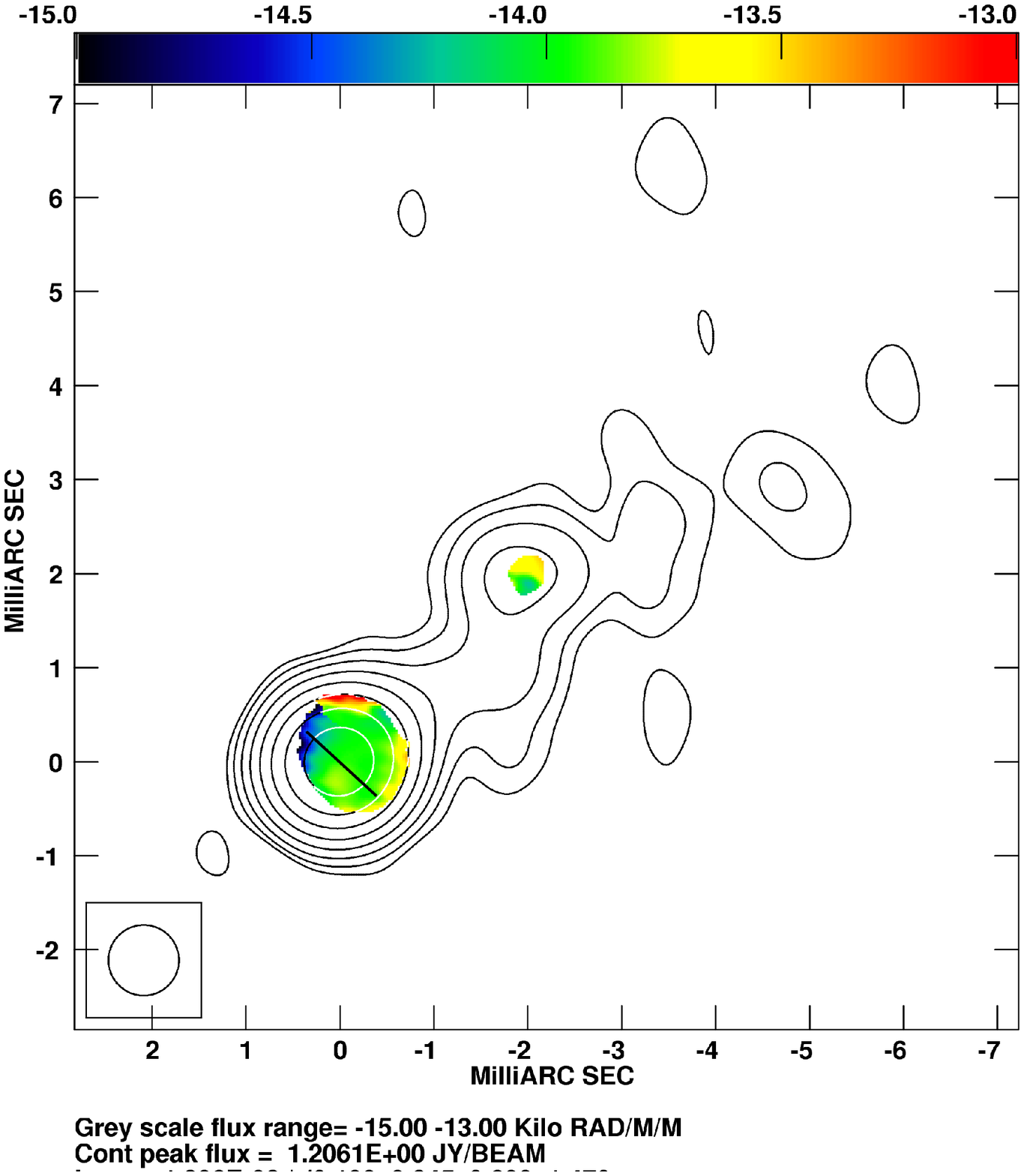}\\

\includegraphics[width=5.5cm, trim=0cm 0cm 0cm 0.9cm, clip=true]{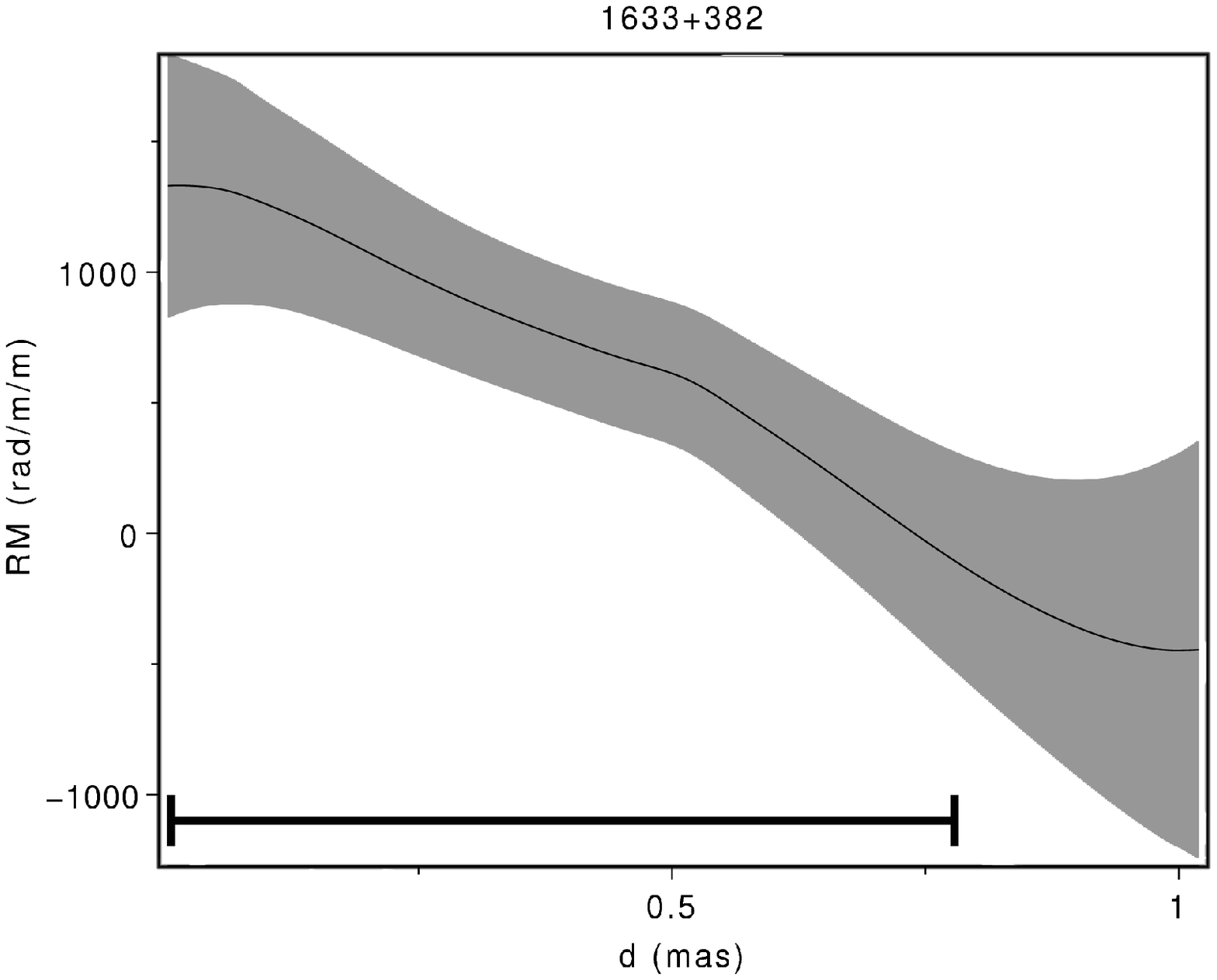}&
\includegraphics[width=5.5cm, trim=0cm 0cm 0cm 0.9cm, clip=true]{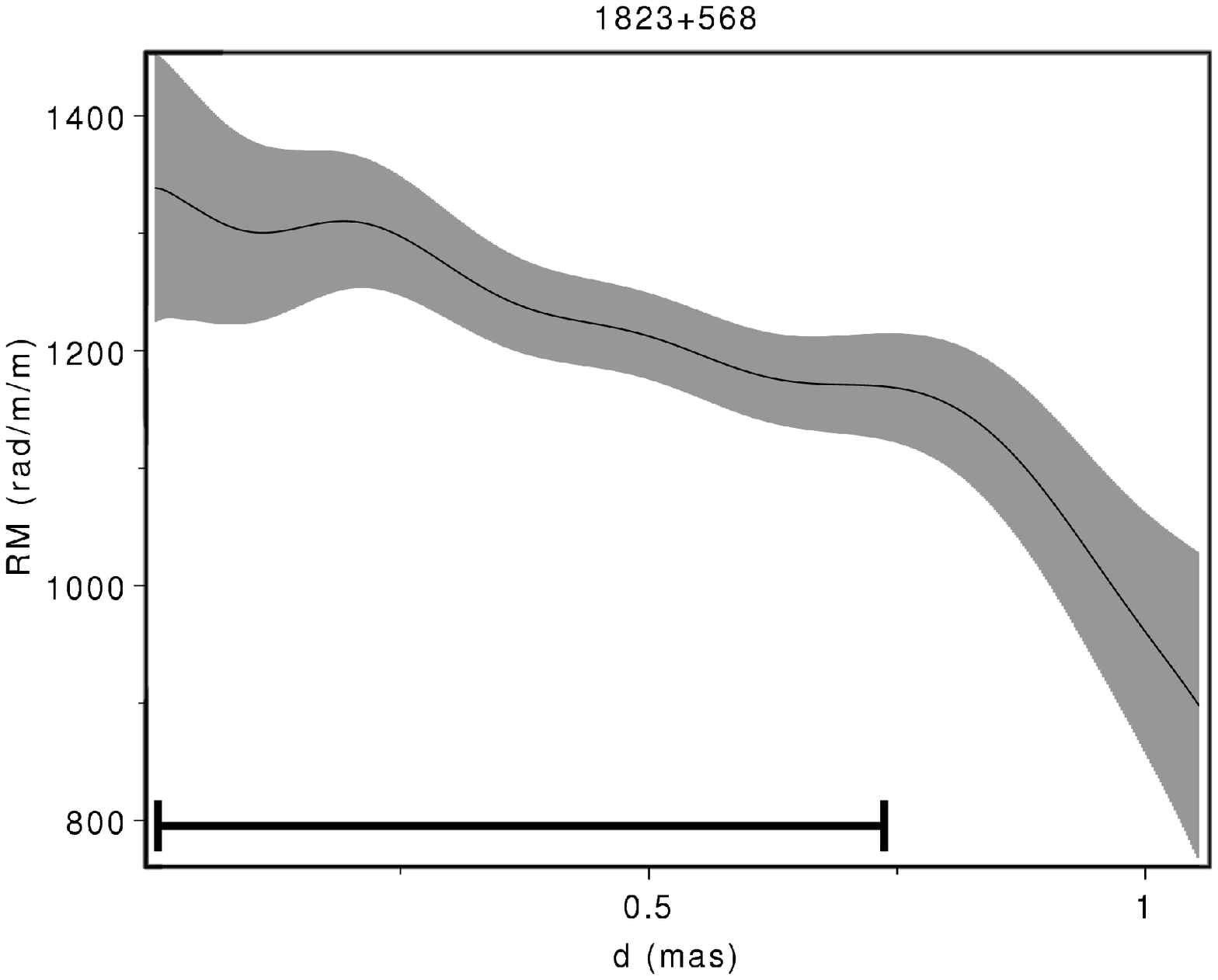}&
\includegraphics[width=5.5cm, trim=0cm 0cm 0cm 0.9cm, clip=true]{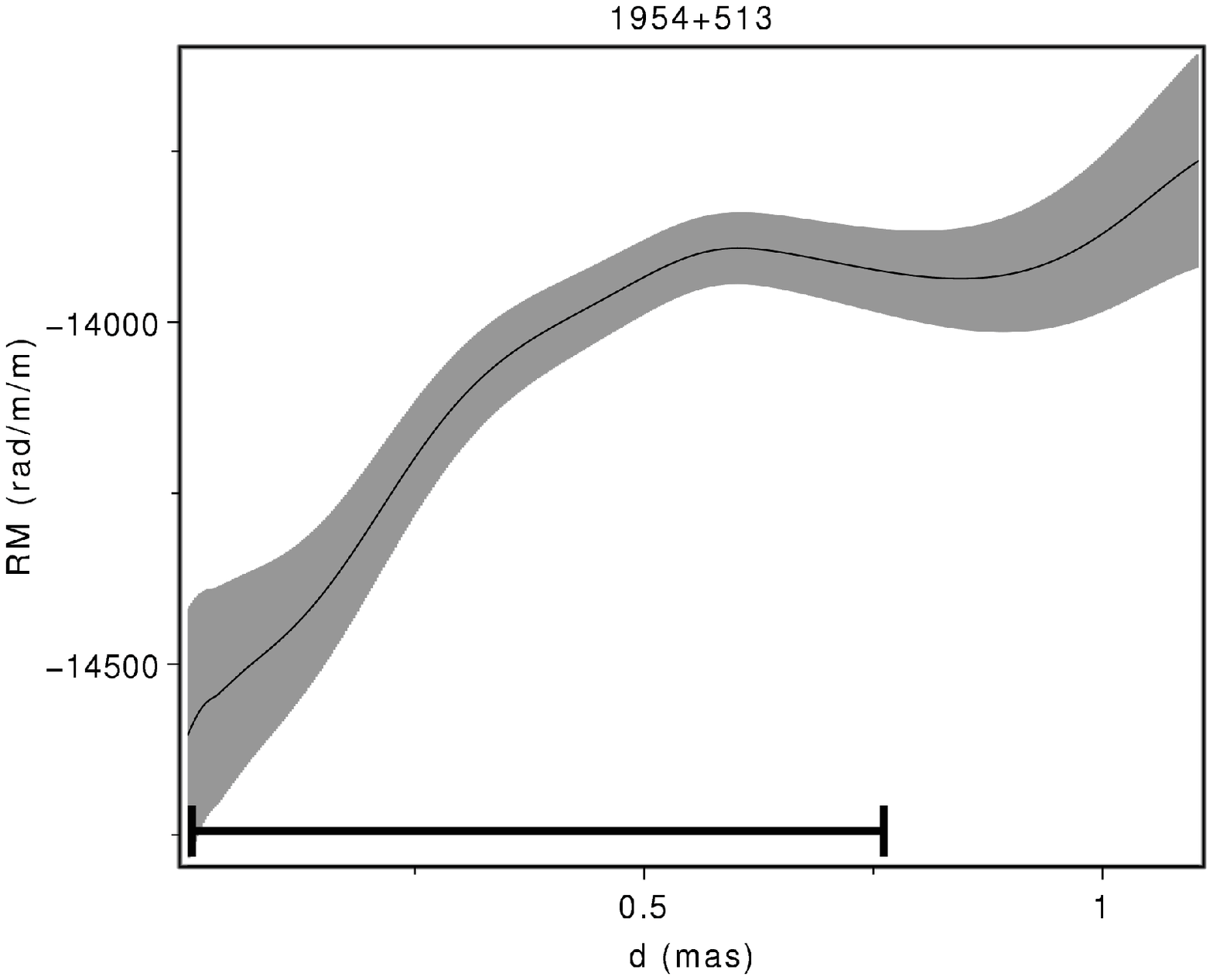}\\

(d) 1633+382&(e) 1823+568&(f) 1954+513\\

  \end{tabular}
\caption{Rotation measure maps. For each of the sources, top:  $RM$ map; bottom: slice of the $RM$ along the thick black line taken on the $RM$ map. Color figures can be found in the online version.}
\end{figure*}

\addtocounter{figure}{-1}
\begin{figure*}
  \centering
  \begin{tabular}{ccc}
\includegraphics[width=5.5cm]{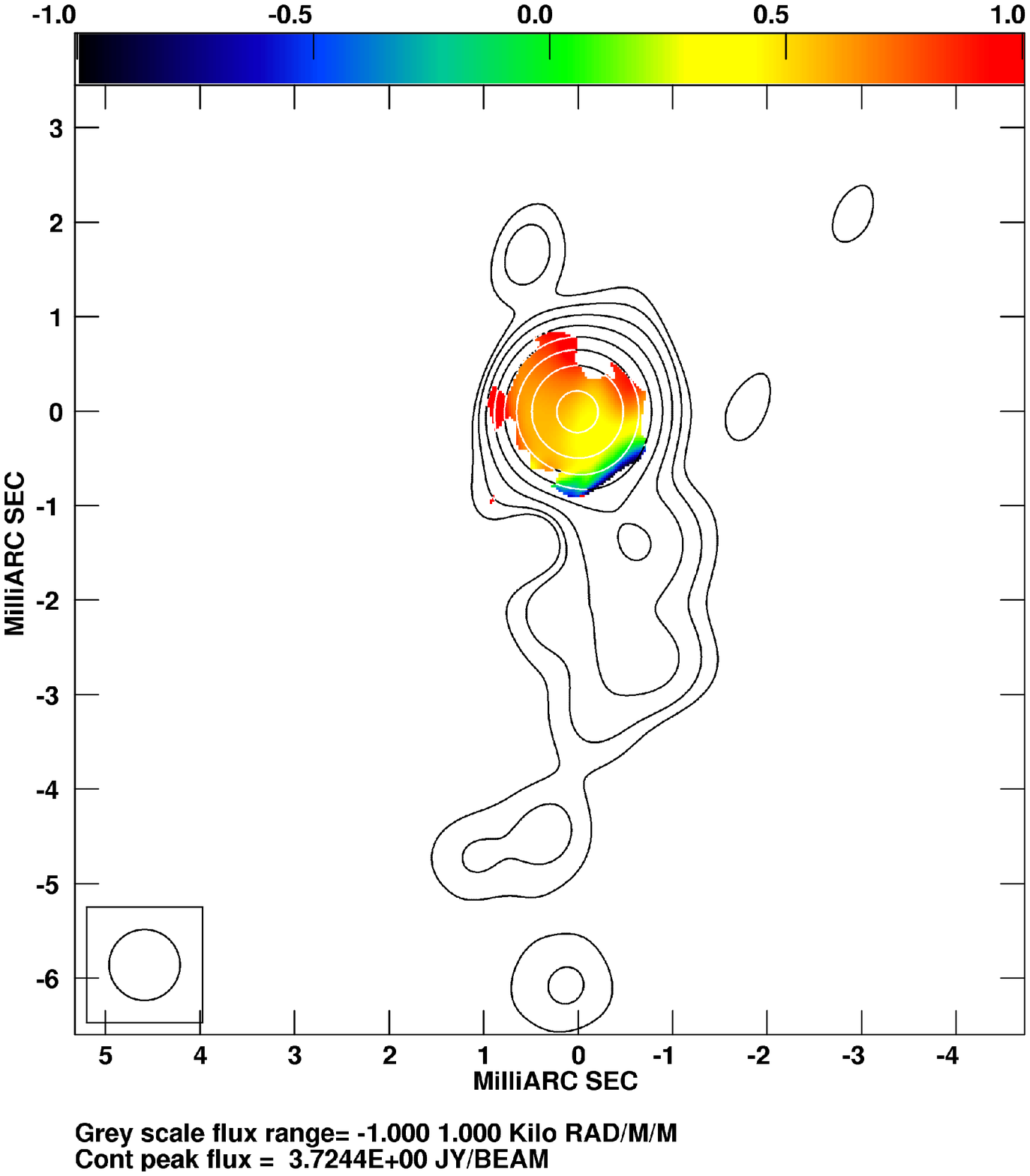}&
\includegraphics[width=5.5cm]{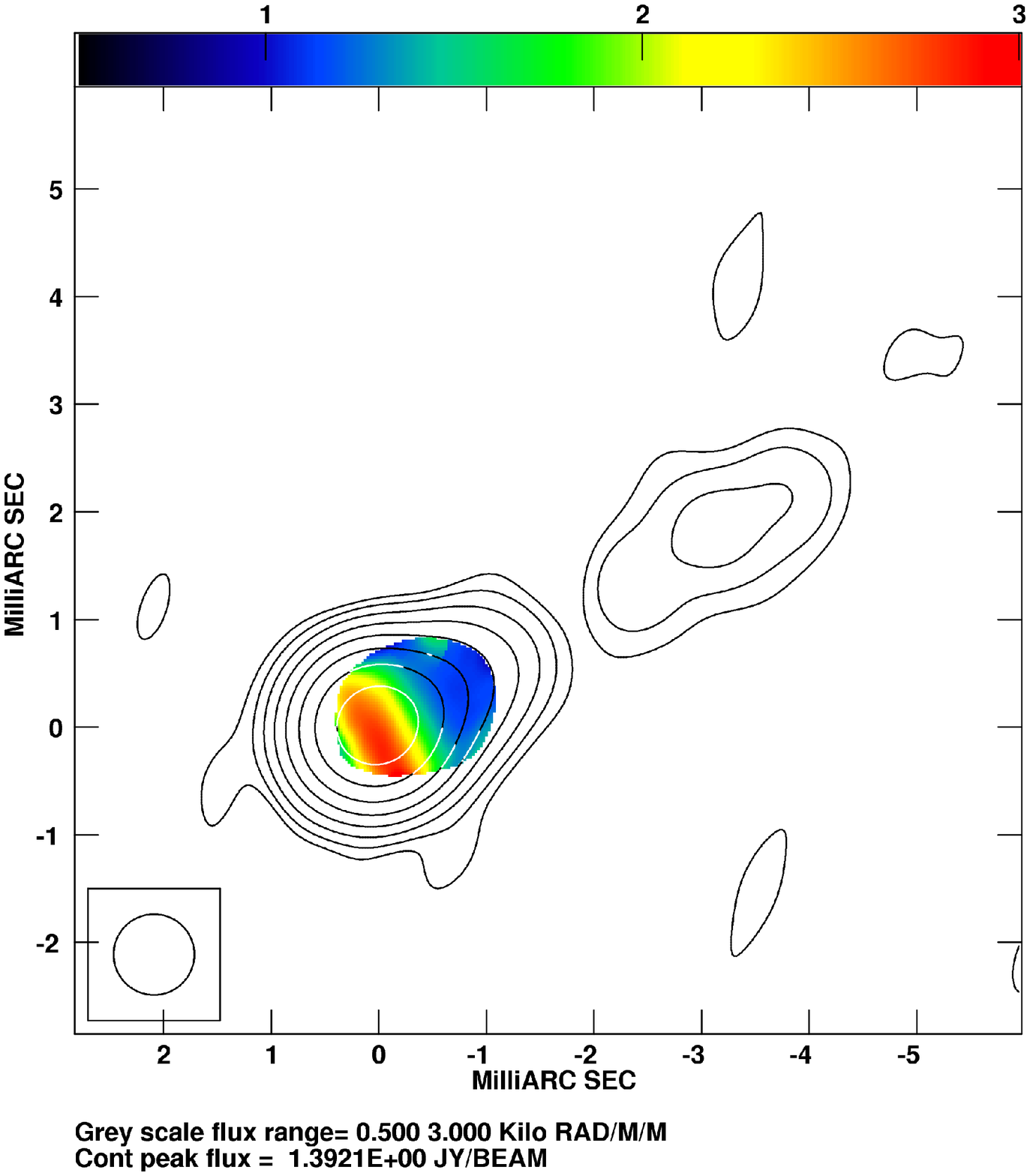}\\

(g) 0420-014&(h) 0745+241\\

  \end{tabular}
\caption{Rotation measures (cont.)}
\end{figure*}

\begin{table*}
\label{contours}
\begin{center}
\caption{Parameters used for the maps.}
\begin{tabular}{c c c c c c}
\hline\hline
Source & $I_{Peak}$ & Bottom Contour & Polarization Sticks & Frequencies used for $RM$ & $RM$ range \\
 &  (Jy/beam) & (mJy/beam) & (mJy/beam)/mas& (GHz)& (rad/m$^2$)\\
$[1]$&[2]&[3]&[4]&[5]&[6]\\
\hline
0133+476& 3.19  & 3.2 &50      &22, 24, 43 &  -4000 -- 1000\\
0256+075& 0.32  & 1.0  &50    &12, 15, 22, 24 & 1000 -- 2000\\
0420--014& 3.72 & 5.6   &200   &12, 15, 22, 24, 43 & -1000 -- 1000\\
0745+241 & 1.39 & 1.4   &40   &12, 15, 22, 24, 43 & 500 -- 3000 \\
0906+430 & 0.69 & 1.4   &10   &12, 15, 22, 24 &-2000 -- 2000 \\	
1633+382 & 2.18 & 2.2   &20   &12, 15, 22, 24 & -500 -- 1000 \\
1823+568 & 1.18 & 2.4   &100   &12, 15, 22, 24, 43 & 500 -- 1500\\
1954+513 & 1.21 & 1.2 &40     &12, 15, 22, 24, 43 & -15000 -- -13000 \\
\hline
\end{tabular}
\end{center}
\end{table*}


Rotation measure is typically found in the VLBA core region; that is, close to the base of the jet. In 0745+241 some $RM$ extending a bit further from the core is found, and only in 0906+430 the $RM$ extends continuously from the core up to a distance of 2 mas into the jet. In 1633+382 and 1954+513 we also find $RM$ in the knots located at 4 and 3.5 mas from the core, respectively. Typical $RM$ values are of the order of some krad/m$^2$, with 1633+382 and 1954+513 showing the most extreme values.

In order to investigate the observed RMs we made slices to analyze the rotation measure across the jet. Plots of the rotation measure across the jet are shown below the $RM$ maps in Figure 1, indicating the region taken for the slice with a thick black line in the map (note that for 0420-014 and 0745+241 no slices are shown, as these sources do not present any indications of variations of the $RM$ across the jet). In all cases, the direction for the slice is taken to be from left to right and, in case of confusion (1633+382, for example), from top to bottom. The horizontal axis of the rotation measure slice plot has been zeroed at the beginning of the slice. {A line in the bottom of the slice indicates the size of the beam. Note that, for the case of 0133+476, the slice is smaller than the beam size. We discuss the significance of these gradients following the criteria given in the Appendix.

\textbf{0133+476} is a quasar showing the $RM$ in the boundaries of the VLBI core. The values obtained for the $RM$ span from -4000 to 1000 rad/m$^2$. Although the slice taken shows a very clear gradient  with a large SNR ($\sigma\sim15$), the region where a possible gradient is seen is very limited, less than one beam width, making this very ambiguous. According to the simulations performed in the Appendix, and together with the fact that it is located in the core, this is not a reliable gradient.

\textbf{0256+075} {shows a gradient in the core region spanning a bit more than a single beam size and forming an angle with the direction of the jet. The values for the $RM$ found in this source vary monotonically from 2000 rad/m$^2$ in the upper region of the core to about 1000 rad/m$^2$ as we approach the upstream region of the jet. However, this gradient is too small both in size and slope to be considered reliable. }

\textbf{0420--014} shows no indications of variations transverse to the jet but does show the $RM$ monotonically decreasing from positive ($\sim+1000$rad/m$^2$) to negative ($\sim-1000$ rad/m$^2$) values as we move further from the central region.
\textbf{0745+241} shows a behavior similar to these of 0420--014 with higher $RM$ values ($\sim$3000 rad/m$^2$) closer to the core and a monotonic decrease to $\sim$1000 rad/m$^2$ at a distance of $\sim$1 mas. 

\textbf{0906+430} has an almost null $RM$ in the core, whereas it takes positive values (+2000 rad/m$^2$) on one side of the jet, decreasing gradually and changing sign down to -2000 rad/m$^2$ on the other side of the jet. This is the source which shows the clearest $RM$ gradient spanning two beam widths along the jet. The slope is about 3 times the typical error and its profile is very smooth, in agreement with all the requirements discussed. Also, the gradient is clearly visible at around one beam size away from the core, in agreement with simulations performed by \cite{Broderick2010}. We consider this $RM$ gradient to be robust.

\textbf{1633+382} Shows a very high (RM=22 krad/m$^2$), featureless $RM$ in the core, but displays an interesting gradient in the knot located at about 4 mas from the core. We discuss reliability of this gradient in section 4.3.2. We note that in Figure 1d we have set the $RM$ range to enhance the $RM$ features in the knot for clarity in the discussion. As high $RM$ values are not included in the scale, the map is saturated for the values occurring in the core.

\textbf{1823+568} shows a very flat rotation measure in the core, with values of $\sim$1200 rad/m$^2$. As we move away from the VLBI core some indications of a transverse gradient taking only positive values arise. This is in general agreement with the idea that finite--beam effects blend different regions and make the detection of reliable features of the $RM$ in the core impossible.

\textbf{1954+513 }shows indications of a gradient across the core region that seem to be faded out (or inverted) as we move less than 1 beam size away. This case is similar to one of the simulations of \cite{Broderick2010}, where spurious rotation measures and sign changes occur, and thus, we do not consider this gradient reliable. 

In general, we find rotation measures to be on the order of several krad/m$^2$, the core of 1633+382 being the most extreme case. Variations of the $RM$ in the optically thick core are, in general, not reliable and give rise to unrealistic gradients \citep{Broderick2010} and \cite{Taylor10}. Thus, such $RM$ variations will not be discussed. 

\section{Discussion}

\subsection{Comparison with previous results}

The source 0256+075 has been previously studied in \cite{Mahmud09a}, where they claimed to have found an $RM$ gradient in the VLBA core, with a gradient reversal that they explained using magnetic tower models proposed by \cite{LyndenBell94} and \cite{LyndenBell96}. However, their gradient was found very close to the core and with a width not significantly exceeding the beam size. Thus, it is likely that the gradients they observe are also due to $RM$ fluctuations in the core.

0420--014 was one of the sources for which an $RM$ analysis was done usin the MOJAVE program. Almost no $RM$ was found by \cite{Hovatta12} for this source in October 2006, with only small patchy traces within one beam size from the core. The difference in the $RM$ detection could be due to the increase of both polarization and the degree of polarization by a factor of 2 in the epoch we study here, allowing us to obtain a better SNR with a similar sensitivity.

\cite{VenturiTaylor} performed observations of 0906+430 in November 1996 and also found clear $RM$ over a long distance from the core in this source. The rotation measure they found in the core is on the order of 100 rad/m$^2$, about one order of magnitude smaller than the results presented here. No $RM$ of this source is shown in the Faraday rotation studies of \cite{Hovatta12}. 

Rotation measure in the sources 0133+476 and 1823+568, and 0745+241 was studied by  \cite{Hovatta12} and \cite{Gabuzda04} respectively. The $RM$ features and sign are different when compared with our work. Although this could be due to some time variation, given their location (optically thick core) and extension (generally, less than 1.5 beam sizes), it is also likely that they might have a non--negligible contribution from spurious $RM$ due to noise in the data.

\subsection{Increase of Core $RM$ with Frequency}
As we observe at higher frequencies, the core $RM$ is expected to increase. This is due to the apparent change in location of the central engine with frequency (i.e, core shift) because of optical depth effects. Thus, observations of the VLBI core at higher frequencies probe regions closer to the central engine, where both magnetic field and particle densities increase. Hence, the core $RM$ is expected to increase with frequency following the formula $RM\propto \nu^a$ \citep{Jorstad07}.  Theoretical estimations provide $a=2$ under the assumption of a toroidally dominated magnetic field and electron density in equipartition scaling as $B\propto d^{-1}$ and $n_e \propto d^{-2}$, respectively, where $d$ is the distance from the central engine, and the outflow is a spherical or conical wind. This value is also in general agreement with estimations for $a$ on other similar sources by \cite{Jorstad07} and \cite{O'Sullivan09}.

We have compared our measurements with previous core $RM$ results in the literature at other frequencies. For our data, we  used the core $RM$ obtained in \cite{Algaba12}. For data from the literature, we used the author's fitted data for the core. If unavailable, we estimated the core $RM$ based on their $RM$ map. We note that, for the case of 0420-014, \cite{Hovatta12} were not able to find $RM$ in the VLBI core and thus we have used the value they found upstream in the jet as the lower limit. To our knowledge, there is no previous VLBI $RM$ data for 1954+513. 

Our results are summarized in Table 2, where column 1 indicates the source; columns 2, 3 and 4 the reference of the lower frequency $RM$ data, the frequency used in the literature and the core $RM$ they found for that frequency, respectively;  columns 5 and 6 the lowest frequency we use from our data to determine the $RM$ and its value; and column 7, the value for $a$ we obtain when we compare previous data with ours at higher frequency.

\begin{table*}
\label{aincrease}
\begin{center}
\caption{RM increase with Frequency}
\begin{tabular}{c c c c c c c}
\hline\hline
Source & Reference & $\nu_1$ & Core $RM (\nu_1) $ & $\nu_2$ & Core $RM (\nu_2)$	 & $a$  \\
 &  & GHz & rad/m$^2$ & GHz & rad/m$^2$ &  \\
$[1]$&[2]&[3]&[4]&[5]&[6]&[7]\\
\hline
0133+476	&H12&	8	&	-216	&	22& -2500	&2.4	$\pm0.4$\\
0256+075	&M09&	4	&	50	&	15& 1530		&2.6	$\pm0.5$\\
0420--014&H12&	8	&	$>548$&	12& 880		&	$<1.2$\\
0745+241	&G04&	5	&	-120	&	12& 2550		&3.5	$\pm0.5$\\
0906+430	&V99&	4.8	&	100	&	12&	-3200 	&3.8	$\pm0.5$\\
1633+382	&H12&	8	&	-235	&	12&	22040	&11.2	$\pm0.4$\\
1823+568	&H12&	8	&	-121	&	12&	1250 	&5.8$\pm0.4$\\
1954+513	&-	&	-	&	-	&	12&	-13800	&-	   \\
\hline
\end{tabular}
\end{center}
\begin{tabular}{l}
\vspace{-0.7cm}
\footnotesize{H12: \cite{Hovatta12}; M09: \cite{Mahmud09a}} \\
\footnotesize{G04: \cite{Gabuzda04}; V99: \cite{VenturiTaylor};}
\end{tabular}
\end{table*}

We find values obtained are around $a=3$, with a lower limit being $a<1.2$ for 0420--014 and the highest value $a=11.2$ for 1633+382. If we exclude these two values, we find the average $\langle a \rangle=3.6\pm1.3$. Except for the case of 1633+382, with exceptionally high $a$, these values are similar to the ones found by \cite{OSullivan09}, ranging from $0.9<a<3.8$, but a bit larger than the ones in \cite{Jorstad07}. The average value found here is not compatible with the theoretical estimation $a=2$ and, indeed, only the value for 0133+476 matches this value to within 1$\sigma$. Two possible explanations for this include  time variations or convolution effects.

Our observations were taken roughly two years later than several of the lower--frequency observations we are comparing with. During this time, several changes may occur in the innermost regions of the jet causing a change of the core RM. Indeed, \cite{ZT01} found changes in the core $RM$ in 3C273 and 3C279 during a period of about 1.5 years. In our case, this is particularly true for  0745+241, 0906+430 or 1823+568, sources that show a change in the sign of the $RM$ for the two epochs. Such $RM$ sign changes have been detected before for 0133+382, 0745+241 and 1633+382 when compared with another set of observations \citep{Algaba12}.


A different reason for our relatively high $a$ value could be that convolved $RM$s are, in general, smaller than the true values and a function of the beam size \citep{Murphy12}. As we observe at higher frequencies, the beam size becomes smaller and, thus, this effect becomes less important. In practice, this implies that the $RM$ observed at lower frequencies was underestimated (or, at least, more so than the $RM$ values studied here), thus producing an artificial increase of the $a$ parameter. However, the range of observed frequencies is relatively small and thus the effects due to deconvolution on the $RM$ are expected to be similarly small.

Another possibility is that the assumptions for the theoretical derivation of $a=2$ are not adequate for these sources. As discussed in \cite{Algaba12}, it is a reasonable assumption to consider that these are in equipartition. However, it is not yet clear how magnetic and electron densities actually scale with distance from the central engine. If these scale in a different way, this could lead to a theoretical prediction for $a$ that is closer to our measured value. For example, if the electron density decays as $n_e \propto d^{-3}$, with the rest of the parameters left unchanged, then the theoretical estimate would be $a=3$. Further investigation of $RM$s at different wavelengths should be performed in order to clarify this.

\subsection{Rotation Measure Gradients}

Although rotation measures have been studied for several decades, it is still not clear when observed variations and gradients of $RM$ are reliable and only recent studies and simulations have dealt with this question. \cite{Taylor10} introduced a series of criteria to discern when a gradient was reliable, \cite{Broderick2010} discussed jet $RM$s from theoretical magneto--hydrodynamic simulations and \cite{Hovatta12} simulated errors of Faraday rotation for their observations.  We discuss these and perform similar simulations in the appendix.

In general, the $RM$s we find are located in the core of the sources. In this region, the $RM$ structure is less reliable and is subject to several blending effects \citep{Broderick2010}. Thus, we will not consider variations of core $RM$. We have found indications of a reliable $RM$ across the jet in two of the sources studied here: 0906+430 and possibly 1633+382. 

The quasar 0906+430 displays a constant $RM$ gradient that spans along the jet up to a distance of 2 mas. At a redshift of $z=0.67$ and its cosmology corrected scale 6.831 pc/mas, this implies a rotation measure gradient extending at least $\sim$14pc. The amount of rotation of the polarization angles that we find is greater than $90^{\circ}$ \citep{Algaba12} and the fractional polarization, although relatively weak ($<1$\% at 12 GHz), is smoothly increasing with frequency. Hence, we can estimate that the contribution of internal Faraday rotation, if any, is negligible. In turn, this indicates that the screen producing the Faraday rotation gradient must be external.

The quasar 1633+382 shows a rotation measure at about 4 mas from the core. If we take into account the errors, the $RM$ runs from around a thousand rad/m$^2$ to zero (or possibly negative values). Unlike the case of 0906+430, this $RM$ is located in a knot of the source, where both total intensity and degree of polarization are enhanced. The range of the $RM$ indicates that, as in 0906+430, it is dominated by external Faraday rotation. 

The case of 1633+382 is indeed close to the limits of what we could consider a reliable gradient: it spans over about 1.5 beams across the jet with a significance $\sigma\sim2$. As we state in the appendix, there is a chance of about $\sim1$\% for this gradient to be a false detection. 
It is possible to have spurious gradients of up to $\sim$230 rad/m$^2$/beam over 1.5 beams (c.f. Figure A4b). This would lead to up to a contribution of $\sim$345 rad/m$^2$ due to the noise, which is about 25\% of the gradient observed in this source. Therefore, the existence of a physical gradient is possible, although we shall discuss it with caution.

$RM$ is also found in the jet of this source in the Faraday studies of \cite{Hovatta12}. Typical values they find in the region we are studying here are $RM\sim200$ rad/m$^2$, which is about five times smaller than the values we find. In their study, \cite{Hovatta12} do not find strong indications of a gradient. If the $RM$ gradient we detect is real, we suggest that it might be due to evolution of the region (see below).

\subsection{Polarization Structure of the Jets}

\begin{figure*}
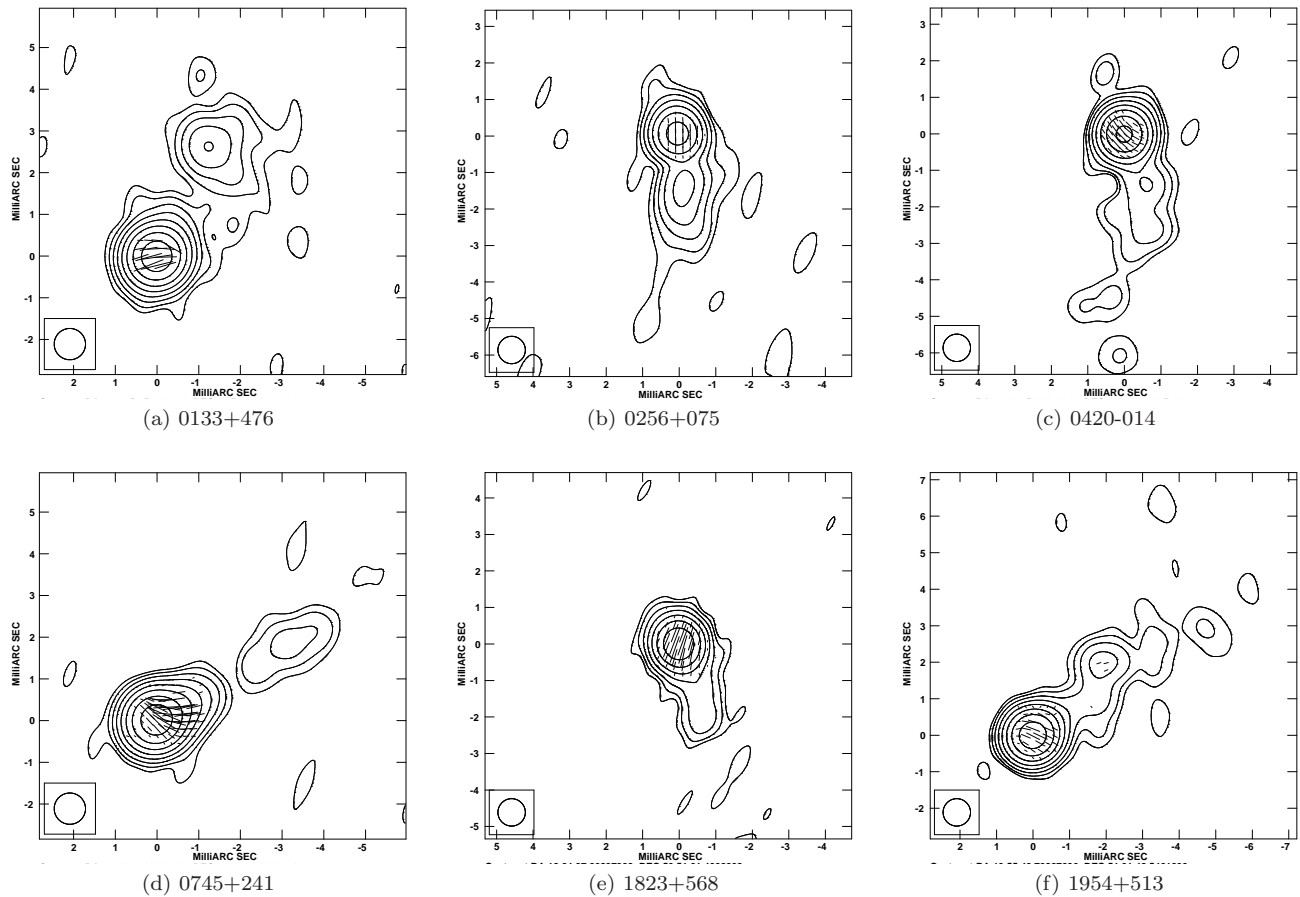

  \centering
  \begin{tabular}{ccc}
\includegraphics[width=5.5cm, trim=0cm 2.99cm 0cm 1.2cm, clip=true]{Efield0133.eps}&
\includegraphics[width=5.5cm, trim=0cm 2.65cm 0cm 1.2cm, clip=true]{Efield0256.eps}&
\includegraphics[width=5.5cm, trim=0cm 2.99cm 0cm 1.2cm, clip=true]{Efield0420.eps}\\
(a) 0133+476&(b) 0256+075&(c) 0420-014
\vspace{0.5cm}\\
\includegraphics[width=5.5cm, trim=0cm 2.99cm 0cm 1.2cm, clip=true]{Efield0745.eps}&
\includegraphics[width=5.5cm, trim=0cm 2.99cm 0cm 1.2cm, clip=true]{Efield1823.eps}&
\includegraphics[width=5.5cm, trim=0cm 2.99cm 0cm 1.2cm, clip=true]{Efield1954.eps}\\
(d) 0745+241&(e) 1823+568&(f) 1954+513\\

  \end{tabular}
\caption{Faraday corrected magnetic field maps for 
sources with $RM$ found only in the core. Contours indicate total intensity (as in Fig. 1), whereas sticks length and direction indicate strength and direction of \textbf{E}--field (see Table 1, column 4).}
\end{figure*}

Once we have obtained $RM$ maps, it is possible to correct for Faraday effects on these regions. We show Faraday corrected electric field maps in Figures 2--4. Here, contours indicate total intensity at 12 GHz as in Figure 1 and direction and longitude of the sticks indicate the direction of the Faraday corrected \textbf{E}--field direction and polarization intensity respectively. Note that we do not show the \textbf{E}--field for these regions where we did not find $RM$, as we are unable to correct it for Faraday rotation. 
A longitude of 1 mas in the polarization sticks corresponds to the number of mJy/beam indicated in column  4 of Table 1.

At a first glance, it seems that variation in the direction of the electric field in the core region is smaller and smoother when corrected for Faraday effects. This is particularly true for the sources 0256+075, 0420--014 and 1823+568. However, in sources such as 0133+476, 0745+241 or 0906+430 there appear to be two differentiated components in the direction of the \textbf{E}--field: one in the VLBI core and one upstream in the jet. We suggest that this second component is caused by an unresolved feature in the jet.

In several sources, the electric field in the core seems to be either perpendicular (0745+241, 1954+513) or parallel (0133+476, 0256+075, 0420--014, 0906+430) to the direction of the jet. This bimodal configuration has been previously observed \citep[see e.g.][and references therein]{Gabuzda2000} and can be interpreted as reflecting the presence of helical magnetic fields associated with the jets. We observe polarized emission from both the front and the rear of the emitting regions but the polarization angles will have a 180\degr offset and, when both contributions are vectorially added, the result will depend on the pitch angle of the jet. With a small pitch angle, the azimuthal component of the magnetic field dominates and the vectorial addition will produce polarization perpendicular to the jet; whereas with a large pitch angle, the toroidal component dominates and the polarization will be parallel to the jet \citep{Asada02}.

In only two sources (1633+382 and 1823+568) there is an evident misalignment between the \textbf{E}--field and jet directions. One of the explanations for this, as above, is the presence of an unresolved (in this case even in polarization) component. Other possibilities include opacity effects on the magnetic field direction or differential Doppler boosting, which would twist the electric vector pattern \citep{Roberts12}. Oblique shocks with arbitrary angles can also cause arbitrary observed polarization angles. 

Due to sensitivity limits, we have been able to detect significant polarization at high frequencies and $RM$ in the jets of only two sources: 0906+430 and 1633+382. We study these two cases below.

\subsubsection{0906+430}

Figure 3 shows the polarization structure of this source. The top panel shows the polarization map as detailed above. The bottom panel shows the degree of polarization $m$ of the map at 12 GHz across the width of the jet in the direction of the $RM$ slice shown in Figure 1. The axis has been zeroed at the centre of the jet. In the degree of polarization slice, areas shaded in grey indicate the regions with polarized flux density below 3$\times$RMS in the polarization images.


There seem to be two regions with different electric field directions along the jet. Close to the VLBI core, the \textbf{E}--field appears to be forming an angle with the direction of the jet, whereas further downstream it becomes perpendicular to the direction of the jet, as in the core. As discussed above, this might be caused by an unresolved component, opacity or relativistic effects. If we investigate the 43 GHz intensity map, we can identify a resolved component located at $\sim1$ mas from the core, and hence we think this might make a significant contribution to the change of the polarization direction.

There are previous studies \citep[see e.g.][]{Marscher10} where a rotation of the optical polarization angle has been seen, presumably due to a structure following a spiral path through a helical magnetic field where the flow accelerates. One of the possibilities is that we are observing a ``snapshot'' of the radio counterpart here, with this structure rotating the polarization angle or the region around it.

When we study a cut of the degree of polarization $m$ along the direction of the gradient (i.e, across the jet), we find a minimum in the centre of the jet and indications of an increase as we move toward the edges, although the errors become rapidly large and, after 0.5 mas from the jet centre, the errors dominate and the polarization drops below 3$\times$RMS. In the region of the $RM$ gradient, even if we take into account the errors, $m$ seems to have a concave shape across the jet.

Overall, one possible way to explain the structure of 0906+430 could be as follows: 
a new component (only resolved at 43 GHz in total intensity) is emerging from the VLBI core and following the path along the intrinsic helical magnetic field, thus altering the direction of the observed polarization closer than 1 mas to the core. Further away, we find three aspects (RM gradient, spine structure of the polarization and shape of $m$ across the jet) that point toward a helical shape of the magnetic field. We note \citep{Contopoulos09} that a spine--sheath polarization structure is not a necessary condition to pinpoint a helical magnetic field, as the observed structure depends on parameters such as viewing and/or pitch angles.

\begin{figure}\vspace{0.4cm}
\begin{center}
\includegraphics[width=8cm]{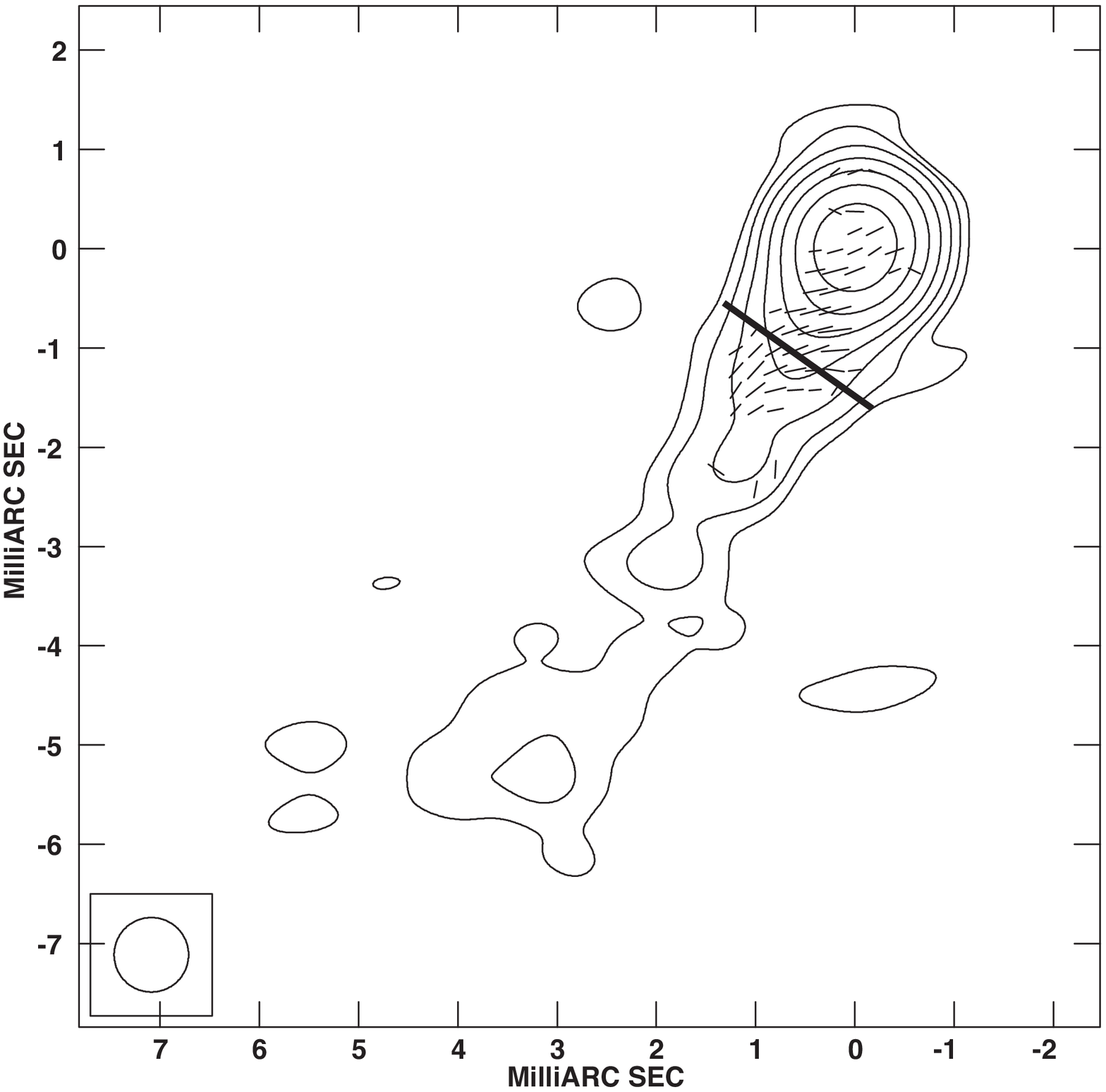}\\
\vspace{-1cm}
\includegraphics[width=8cm,trim=0cm 0cm 0cm 1.8cm, clip=true]{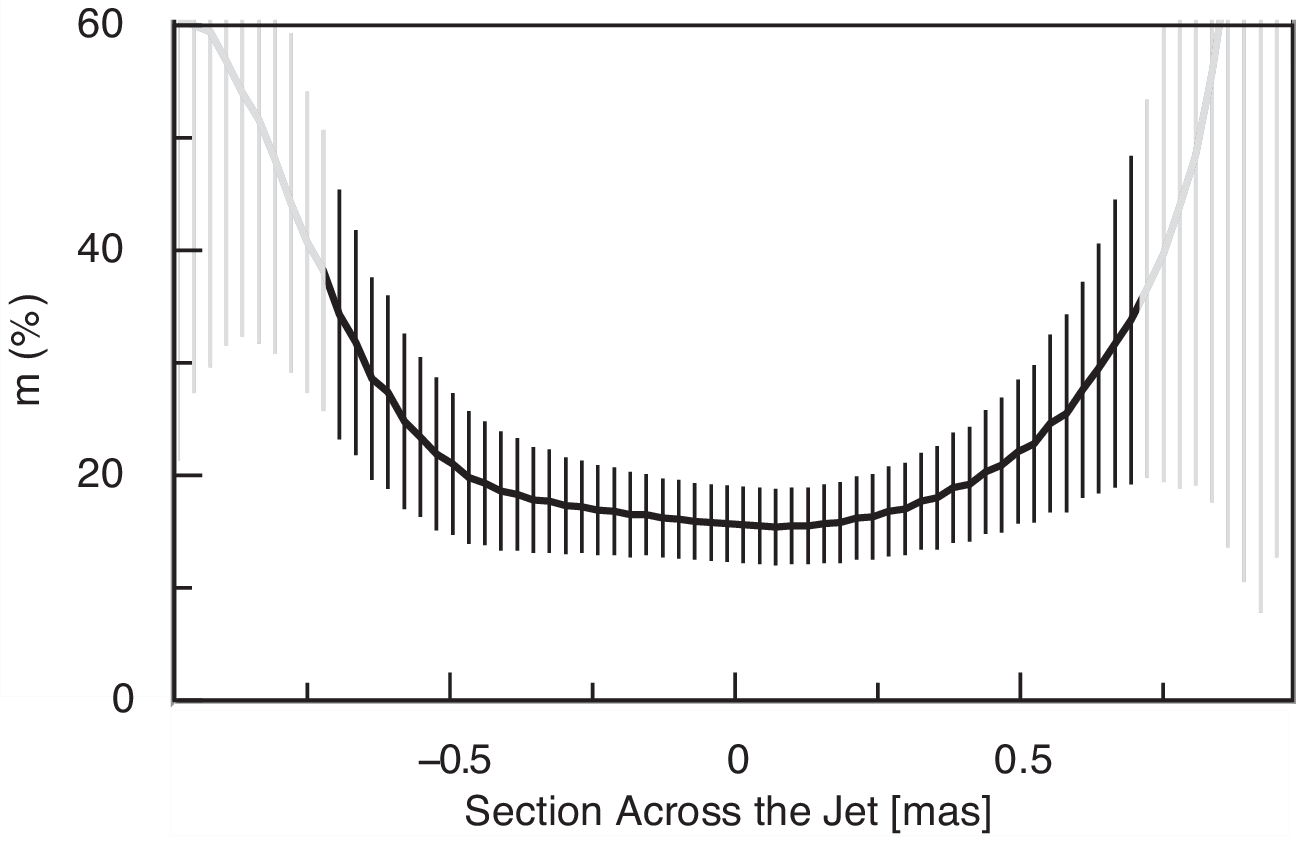}\\
\end{center}
\caption{Polarization of 0906+430. Top: polarization map. Bottom: degree of polarization across the jet following the direction of the $RM$ slice. Grey indicates the areas with polarized flux density below 3$\times$RMS in the polarization images.}
\end{figure}

\begin{figure}\vspace{0.5cm}
\begin{center}
\includegraphics[width=8cm]{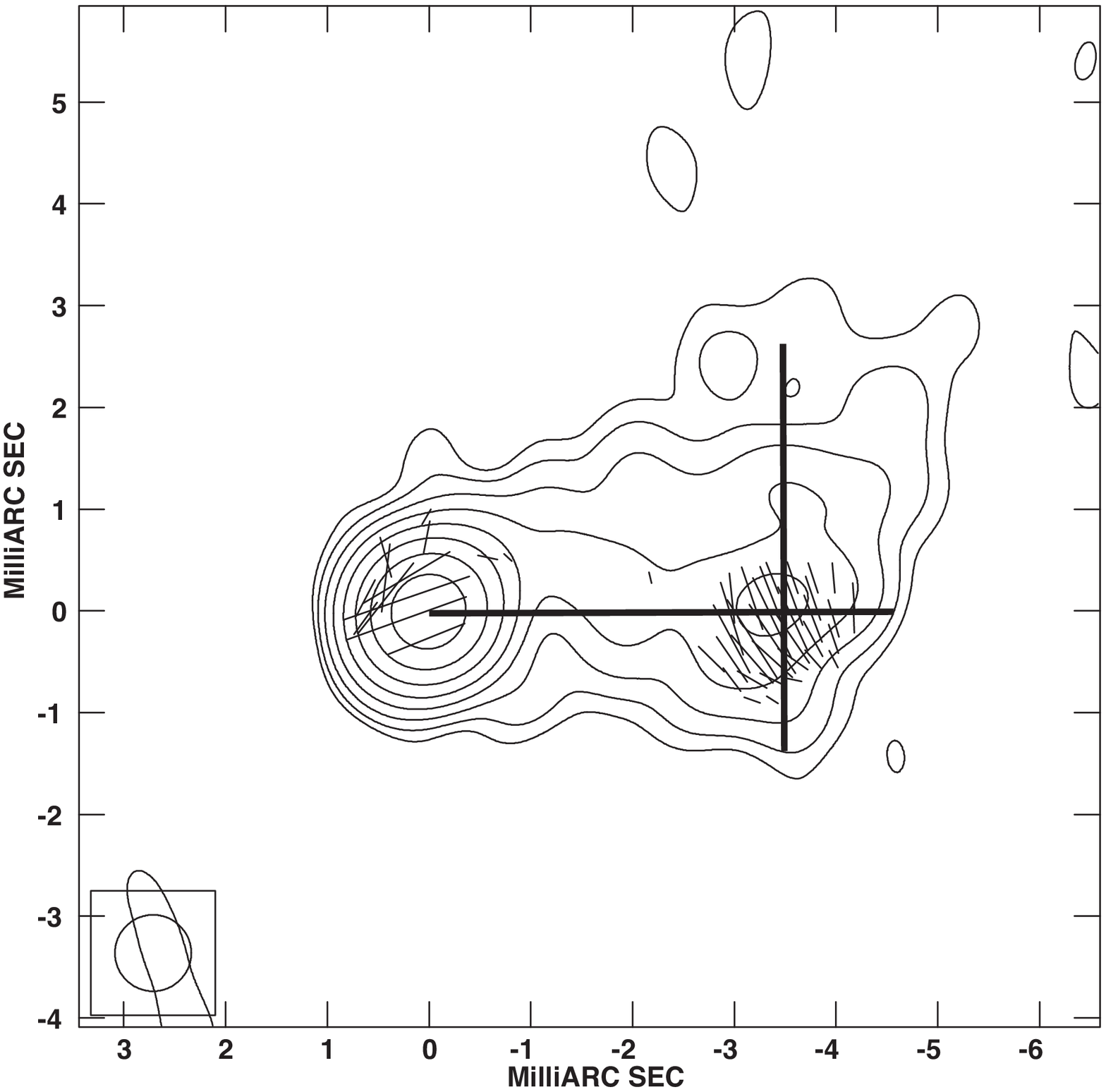}\\  \vspace{-1cm}
\includegraphics[width=9cm,trim=0cm 0cm 0cm 1.8cm, clip=true]{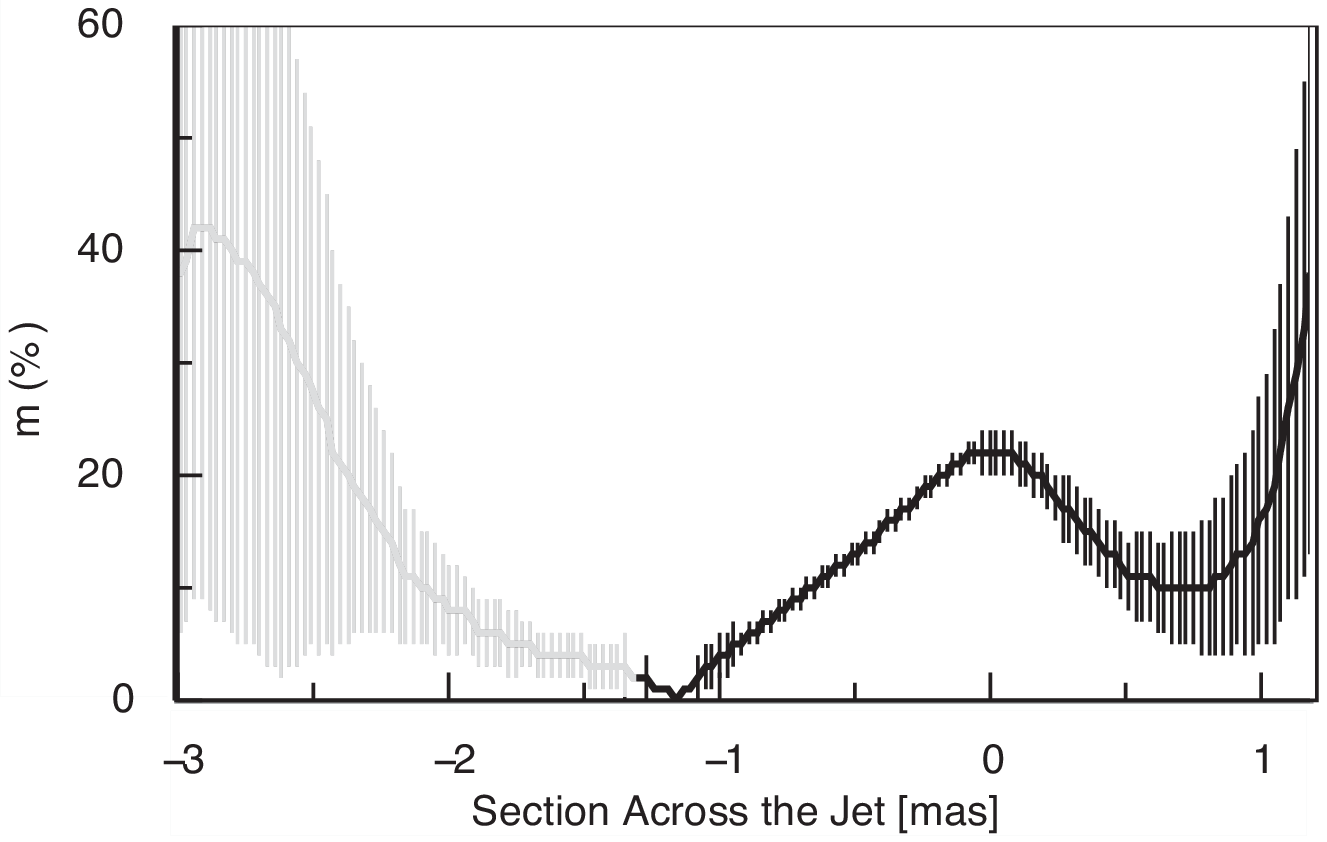}\\  \vspace{-0.5cm}
\includegraphics[width=9cm,trim=0cm 0cm 0cm 1.8cm, clip=true]{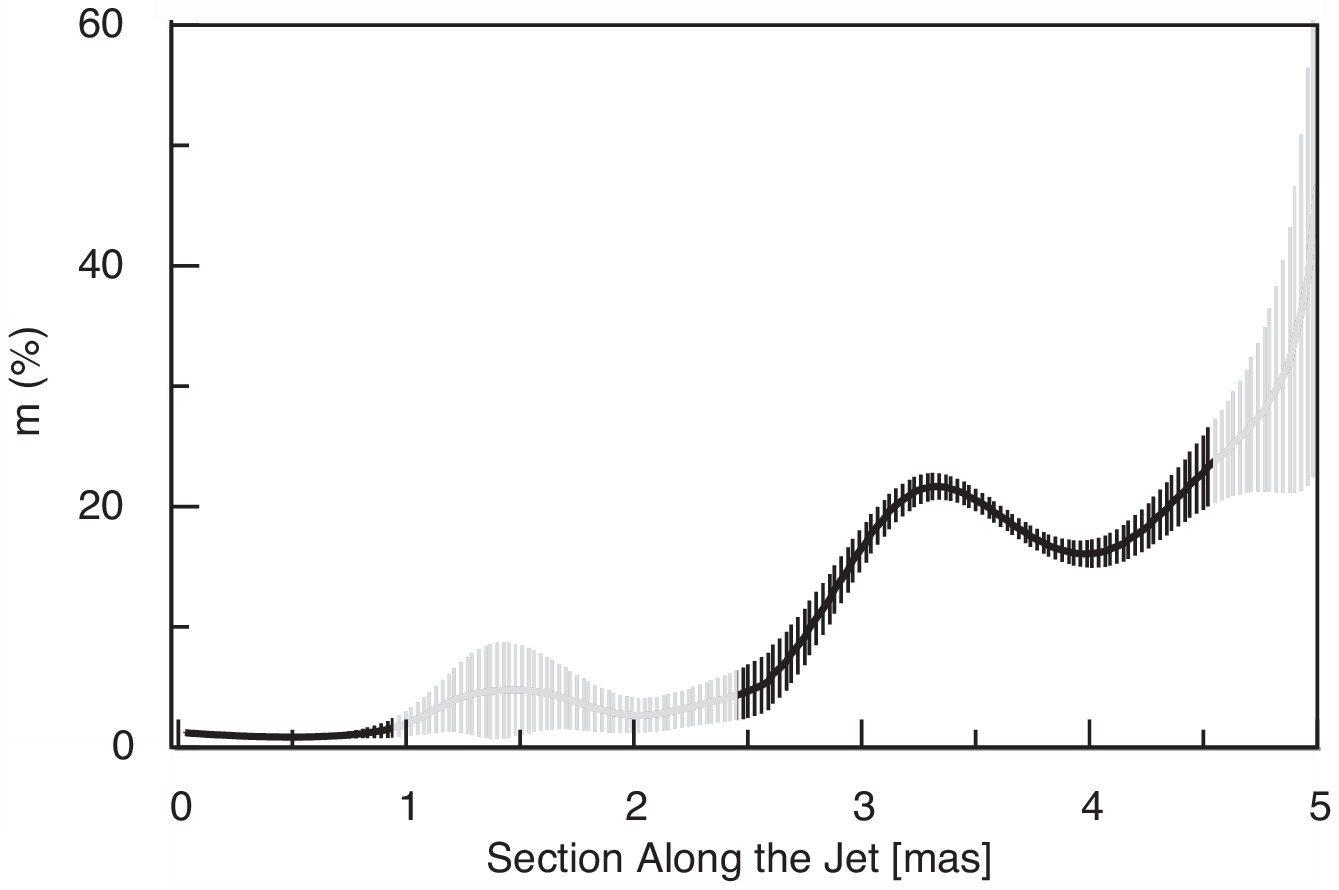}\\
\end{center}
\caption{Polarization of 1633+382. Top: polarization map. Middle: degree of polarization across the jet following the direction of the $RM$ slice. Bottom: degree of polarization along the jet. Grey indicates these areas with polarized flux density below 3$\times$RMS in the polarization images.}
\end{figure}

\subsubsection{1633+382}

Figure 4 shows the polarization structure of this source. The top panel shows the polarization map; the middle panel, the degree of polarization $m$ of the map across the width of the jet in the direction of the $RM$ slice zeroed at the centre of the jet. The bottom panel shows $m$ along the jet zeroed at the core. As before, in the degree of polarization slices, areas shaded in grey indicate the regions with polarized flux density below 3$\times$RMS in the polarization images.

In the region of the jet where we have been able to correct for Faraday rotation, the \textbf{E}--field seems to form an angle with the direction of the jet. When we take a slice perpendicular to the direction of the jet to study the polarization structure of this feature, we find hints of a complex W-shape, although, given the errors both in total and fractional polarization, the only significant feature is a peak of the fractional polarization in this region.

A smooth sheath around the jet cannot produce a region of enhanced polarization. Hence, we speculate, based on the total intensity and polarization maps, that the reason for this shape is a component arising from the polarization of the knot itself. Indeed, the local peak of the degree of polarization is coincident with the intensity peak of the knot, and both shapes also agree within the errors.  We note that this is not unique, and has also been observed in other sources \citep{Algabaprep}. We also note that magnetic fields do not need to be perpendicular to the direction of the jet to be caused by a shock: oblique shocks with arbitrary angles, leading to different directions of the magnetic field are also possible.

In order to check if such a knot is compatible with the observations, we have obtained a slice of $m$ along the jet. We observe a very low $m$ near the core, but a peak of $m\sim20$\% at a distance $\sim3.5$ mas (corresponding to $\sim$700 pc, deprojected), coincident with the location of the knot. According to \cite{Liu2010} and MOJAVE data, this component has been moving at a velocity of $\sim0.32$ mas yr$^{-1}$, gradually decelerating until reaching its current position around 2003, where it seems to remain steady. 

One possibility is that this is a component that has been gradually decelerated by interaction with the surrounding media and has developed a shock front with compressed magnetic field. Given the degree of polarization, we can conclude that the underlying magnetic field is highly ordered. Compression across the jet will produce an enhancement of the transverse component of the magnetic field. \cite{Algaba12} found that, for this source, the magnetic field at 1pc from the central engine is $B(1pc)=0.7G$. If we assume  $B\propto R^{-1}$, which is in agreement with equipartition arguments and can be used to describe the observed synchrotron emission in compact regions of VLBI jets, we find that the corresponding magnetic field strength in the region of the knot is around $B(1pc)\sim1mG$. This is a relatively high magnetic field strength but is in agreement with the one found in other sources at similar scales \cite[see e.g.][]{Owen89}.

There is another possibility that could also give rise to an intrinsic Faraday--corrected polarization forming an apparent angle with the jet in this region. If we analyze the structure of this source at higher frequencies, a component close to the core ($\sim0.5$ mas) is observed in the north-west direction (see e.g. the 7-mm monitoring by Boston University Blazar Group).This could be an indication that the direction of the jet has actually changed and new components are now being ejected in a direction that is tilted with respect to the previous one. Alternatively, the component located at 0.5 mas may be a standing component \citep{Liu2010}, thus indicating the truly intrinsic direction of the jet and that the actual jet is curved. If this is the case, polarizations both in the core and in the knot might be showing the true direction of the jet in these regions (i.e., the \textbf{E}--field would be parallel to the direction of the jet).

We have obtained new slices of $m$ assuming that the local direction of the jet is the one shown by the \textbf{E}--field vectors and we find that the previous discussion does not significantly change: namely, the W-shape across, and the peak along, the local direction of the jet in the polarization are still found. Under this interpretation, the possible $RM$ gradient found would be \emph{parallel} to the local direction of the jet. This could still be explained in terms of a shock, if we assume that we have a strong compression shock where the \textbf{E}--field is enhanced, followed by a dissipation zone producing the $RM$ gradient. This shock might have been form recently, based on the proper motion of this component, which may explain why such a gradient was not observed in \cite{Hovatta12}, as their observations were done in September 2006, more than two years prior to ours.

However, we are currently unable to distinguish between these scenarios. According to \cite{Hovatta}, the viewing angle for this source is only $2.5^{\circ}$, and so we are looking at it almost head--on, which makes it difficult to adequately study its de--projected features. Also, observations with more sensitivity would be necessary at higher frequencies to find out the actual direction of the jet at different scales. Further observations of the knot studied here will be necessary in order to understand its motion and \textbf{E}--field evolution in the forthcoming years.

\section{Conclusions}

We have obtained polarimetric VLBA observations of 8 AGNs and presented their rotation measure analysis. Faraday rotation is found in the VLBA cores with typical values for the rotation measure on the order of some thousands of rad/m$^2$, except for the sources 1633+382 and 1954+513, with core $RM$s up to 22 krad/m$^2$ and -13 krad/m$^2$.  Due to sensitivity limits, jet $RM$ is detected only in the jets of two sources: 0906+430 and 1633+382.

Core $RM$s found here are larger than in previous studies performed in the same sources at lower frequencies. If we assume a dependence of the form $RM\propto \nu^a$, derived by  \cite{Jorstad07}, we find an average of $a=3.6\pm1.3$, in agreement with estimations made for similar sources \citep{OSullivan09,Jorstad07}. However, most of the individual values are higher than the theoretical estimations that provide $a=2$ under the assumption of a toroidal dominated magnetic field and electron density in equipartition. Different explanations for this include a time variability of the $RM$ (supported in some sources by a sign change over different epochs) or a convolution effect causing $RM$s to be underestimated at lower frequencies.

In order to discuss the reliability of the $RM$s found here we have performed simulations based on our multi--frequency data. Our simulations indicate that for $RM$ variations to be robust, they have to span over at least 1.5 beam sizes with a SNR$>3$, although we have found that these conditions can be \emph{slightly} loosened as stable gradients over a few beam sizes should not arise due to random variations or beam effects. These results agree with previous discussions by \cite{Taylor10} and \cite{Hovatta12}.

The Faraday--corrected direction of the core magnetic field is, in general, either aligned or parallel with the direction of the jet. This can be interpreted as an indication of an underlying helical magnetic field. Two sources seem not to follow this trend. In 1633+382 there were various possibilities including a change of the jet direction and in 1823+382 we propose that the rotation of the polarization angle may be due to an unresolved component. This interpretation is supported by observations at higher frequencies and slices of the degree of polarization, revealing this component.

In some sources (0133+476, 0745+241 and 0906+430) we observe two different regions for the polarization angle. Based on the previous interpretation, we suggest that this is due to the unaltered \textbf{E}--field in the core plus an additional component that, being still unresolved at 12 GHz, has moved away from the core enough to be detectable in polarization.

We find that the $RM$ gradient in the jet of 0906+430 (and possibly 1633+382) is reliable. The combination of the results from the $RM$, the spine polarization structure and the concave shape of the degree of polarization across the jet seem to indicate the presence of a helical magnetic field in 0906+430 beyond 1 mas away from the core. The case for 1633+386 is unclear and different possibilities, such as a compression shock or diffusion of enhanced magnetic field, are given.

\appendix

\section{Reliability of Rotation Measure Gradients}
The subject of rotation measure gradients in the last decade has been a controversial one. Due to the difficulty in addressing the $RM$ errors and the extend of the $RM$ compared with the map beam sizes, the reliability of $RM$ gradients is still debatable. Initially one might think that, when the size of the region where the observed rotation measure is comparable to the beam size, we cannot truly resolve the $RM$ features and the gradients we find might be the spurious product of the integration of several random features within the restoring beam, and not real physical ones. In order to clarify the situation, \cite{Taylor10} summarized a series of criteria in order to establish observational requirements for a reliable $RM$ gradient. We summarize and discuss these requirements here.

As a gradient will exist whenever two values differ, it is clear that, at least, three independent measurements are needed. An initial (but too conservative) guess is that for each element of beam width one can only obtain a single independent value and hence, a gradient spanning along at least three beam widths is needed in order to speculate about a gradient. However, this requirement is too restrictive. The position accuracy of polarized intensity images is below a small fraction of the synthesized beam size, because in the calibration process the same complex gains are applied to both parallel and cross visibilities \citep{Leppanen95}. \cite{Lobanov05} argues that the minimum resolvable size is determined not only by the beam size but also by the signal--to--noise ratio, concluding that, in general, at $SNR>4$, the minimum resolvable size is always smaller than the beam FWHM. For a typical $SNR\sim20$, this implies a resolvable size of $~1/3$ of the beam size; i.e, $\sim0.25$ mas for our case. 

\cite{Murphy12} argue that an $RM$ gradient can change its value when convolved with different beams, but the gradient itself is not destroyed even when the jet is unresolved, although they do not take into account spurious gradients arising from noise. \cite{Hovatta12} perform a series of simulations specifically dealing with this issue that indicate that the gradient is already considered significant with a size of 1.5 beams, provided it exceeds 3$\times$RMS. This is consistent with similar criteria described by \cite{Broderick2010} and in agreement with the discussion by \cite{Lobanov05}. 

To study these aspects in a more quantitative way we followed \cite{Hovatta12} and performed a series of simulations. In their work they used frequencies between 8 and 15 GHz and explicitly pointed out that independent simulations should be done if a different set of frequencies were to be used, as it is the case here. Hence, we used their procedure for our case.

First, the stokes I model of the CLEANed source was used. We then created fiducial Q and U models by setting them to be a known fraction of Stokes I (which leads to a constant EVPA and polarization across the source). We then produced the UV data based on these models and added random independent noise with a normal distribution to the data. Using this method we produced 500 sets of UV data for each frequency, which were used then to derive new Q, U, polarized and respective noise maps. We checked that the resulting maps had noise levels and distributions statistically compatible with the real observations.

We then created simulated $RM$ and $RM$ noise maps using the values of the polarization angles from the simulated 12, 15 and 22 GHz maps, to which we added a random value between $\pm4$\degr to account for EVPA calibration errors. Errors on the $RM$ were obtained by the quadratic sum of different terms, namely the errors from the noise map, estimated from the variance-covariance matrix of the linear fit as the error of the slope, and the simulated propagated errors from the polarization maps. This is done because, although errors from the covariance matrix alone are adequate for each pixel, we still need to take into account that pixels are not independent. We note that we do not include the EVPA calibration errors, as they will have no effect in $RM$ gradients \cite{Mahmud09b}, which we are interested in. Figure A1 shows a sample of a simulated $RM$ map.

Typical error in the $RM$ is about $\sim150$ rad/m$^2$, which is a bit smaller than the error in the real $RM$ in our observation at the same position ($\sim250$ rad/m$^2$). \%Nonetheless, when we carefully check our estimations, we are not able to find a definitive cause for this. 
We estimate that there is a probability of $\sim$5\% to obtain a simulated $RM$ with an error similar or larger than the observed one. Thus, it is possible that our observation is within this case. It might also be possible that this difference arises due to underestimation of the non-thermal errors due to the procedure in creating the Q and U models, or the evaluation of the pixel dependence. After carefully checking out estimations we think this is not the case.

Figure A2 shows the distribution of obtained $RM$ values at a point in the jet for the source 1633+382 close to the area where the actual $RM$ is observed. Values are about a third smaller than the ones in the actual observation, around few hundred rad/m$^2$. The distribution of $RM$ values found is as follows: the average and the most probable values are  -8 rad/m$^2$, and -9 rad/m$^2$, respectively, consistent with a distribution that has no $RM$. The standard deviation of the distribution is 104 rad/m$^2$.

\begin{figure}
  \centering
\includegraphics[scale=0.4, trim=0cm 1.5cm 0cm 0cm, clip=true]{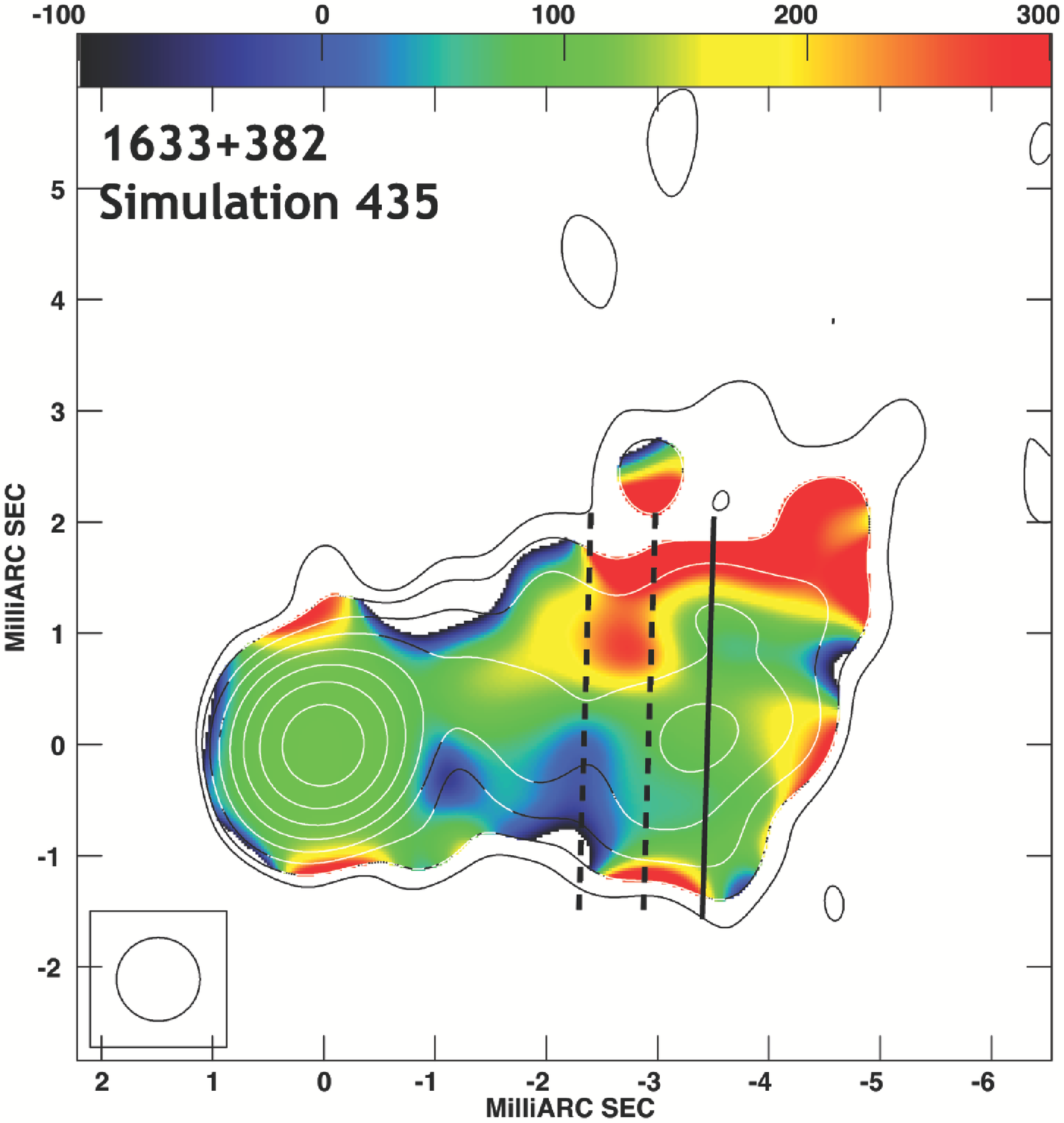}
\renewcommand{\thefigure}{A\arabic{figure}}%
\caption{Slices transverse to the jet in 1633+382 overlaid on a simulated $RM$ map. Straight line indicates the slice used to obtain the simulated $RM$ gradients. Dashed lines indicate slices that were taken to check for the consistency of the method. Color figure can be found in the online version of the journal.}
\end{figure}

\begin{figure}
  \centering
\includegraphics[width=8cm, trim=0.1cm 0.0cm 15cm 10cm, clip=true]{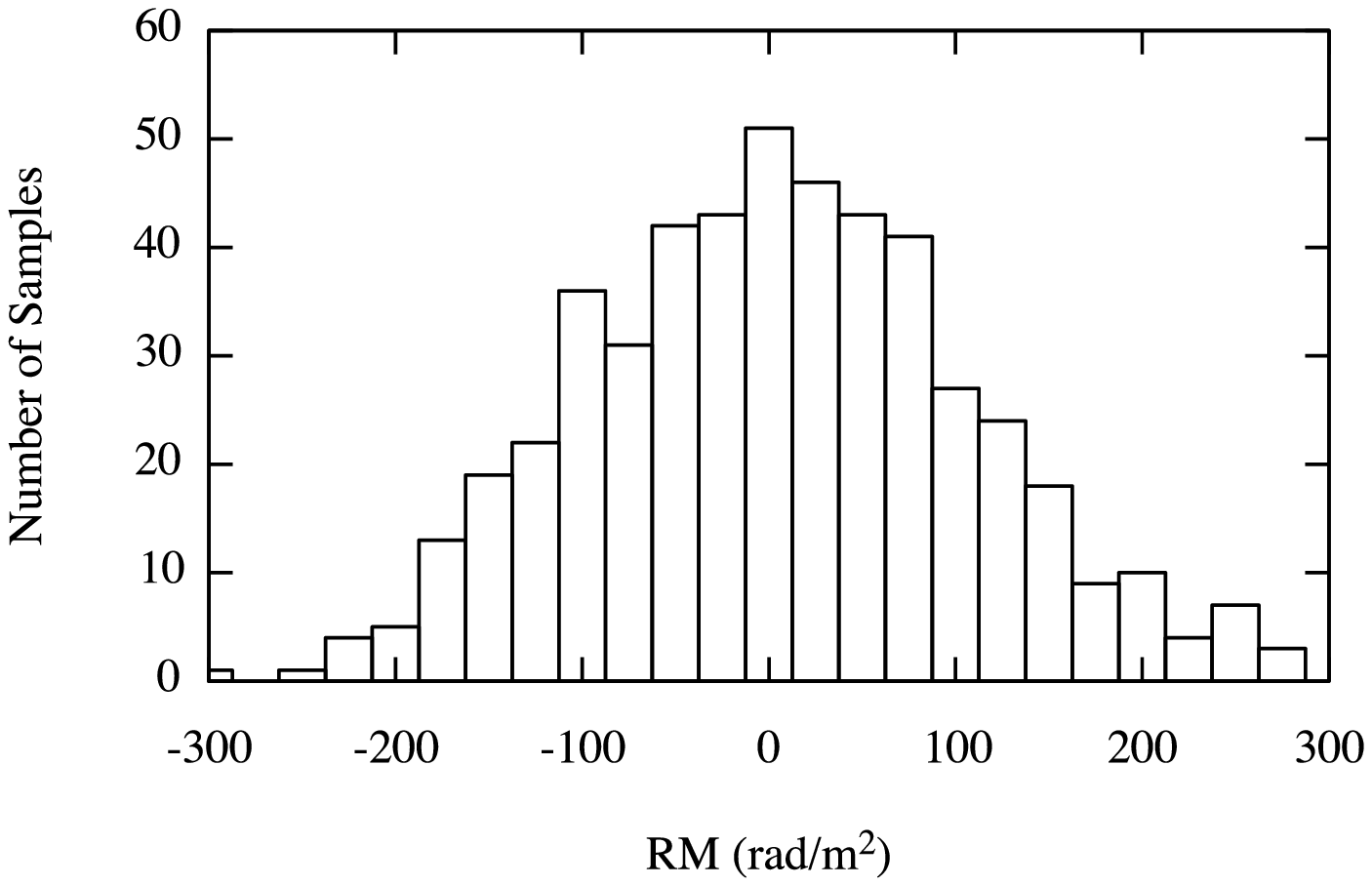}
\renewcommand{\thefigure}{A\arabic{figure}}%
\caption{Distribution of simulated rotation measures in the jet of 1633+382}.
\end{figure}

For each of the simulated $RM$ maps, we took a slice across the jet, as shown in Figure A1 by the straight line. Ideally, we would like to take different slices of various sizes separated by at least one beam width to ensure they are independent. However, given the structure of the source, this is not possible, particularly for slices smaller than $\sim$2.5 beams. In order to overcome this issue we have used the following procedure:

We first set the longitude of a single slice to 1, 1.5, 2, 2.5 and 3 beam sizes respectively} (we note here that 1 beam corresponds to 0.75 mas), all centered around the jet spine. In total, we obtained 2500 slices that were used to calculate simulated $RM$ gradients as follows: we fitted a simple line to the slice. We estimated the gradient error as the maximum between the error in the slope fit and the largest error bar in the $RM$ slice. Because this method does not provide independent measurements, we repeated this process for another two different slices shown by dashed lines in Figure A1 as a consistency check. We find that the results are similar for all three slices.Figure A3 shows the distribution of $RM$ gradients obtained by this method for the straight slice, which is taken along our observed $RM$. In all cases, the simulated $RM$ is centered at $\sim$ 0 rad/m$^2$, as expected, with the standard deviation decreasing as we move to slices larger than 1 beam size.

To study when these $RM$ gradients are significant, we have analyzed their SNR by dividing them by their error, defined as above. Results can be seen in Figure A4, where the number of $RM$ gradients are plotted in terms of their SNR. We note here that none of these $RM$ gradients are real. However, when a slice with the size of only 1 beam is analyzed, there is still a significant number of spurious gradients with $\sigma=$SNR$>3$. The number of these gradients decreases dramatically as we study slices with 1.5 beam sizes or larger: only 4 sources for 1.5 beam sizes and none of them for larger ones had $\sigma>3$ in our simulation.

This is clearly shown in Figure A5, where we plot the number of false positives (i.e, the proportion of sources that exceed 1$\sigma$, 2$\sigma$ and 3$\sigma$ respectively) against the size of the slice. This fraction goes rapidly to zero as we increase the number of error limits  $\sigma$ that are imposed. In general, 1$\sigma$ gradients will not be reliable even for large slices but a level of 2$\sigma$ will be enough when the width of the slices used is at least 2 beams. In most cases, even a $2\sigma$ detection level over 1.5 beams will be reliable, with a chance of finding false gradients in only 0.8\% of these.


Summarizing, the results of our simulations agree with the ones shown in \cite{Hovatta12} and the suggestions by \cite{Taylor10}, although we consider the latter too restrictive. While slices with 3 beam widths are optimal for a reliable detection of a $RM$ gradient, in the most general case smaller slices can also provide a reliable gradient, provided the error treatment and SNR is adequate.

\begin{figure*}
  \centering
  \begin{tabular}{ccc}
\includegraphics[width=5.5cm, trim=0cm 0.0cm 15.4cm 10cm, clip=true]{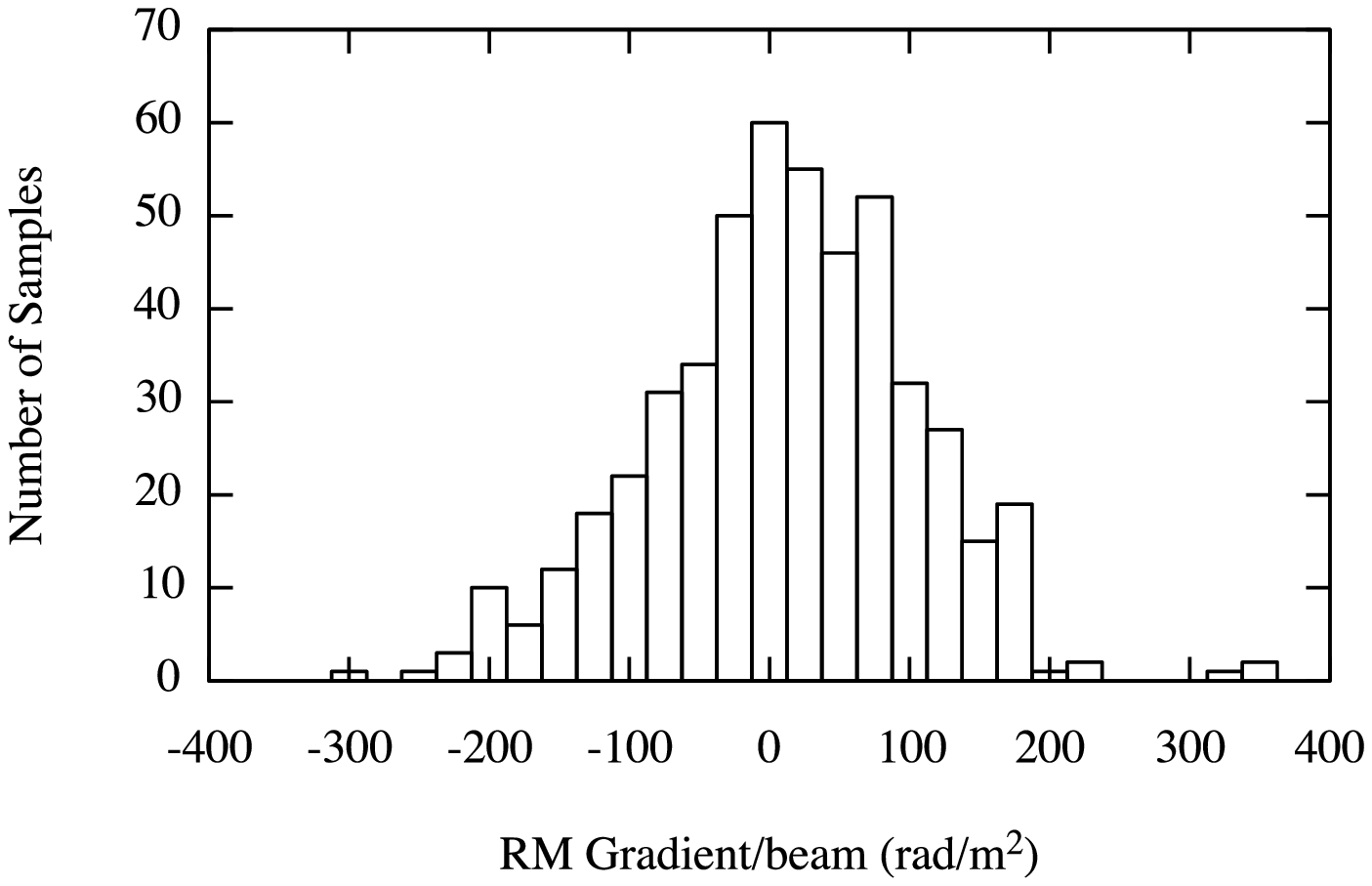}&
\includegraphics[width=5.5cm, trim=0cm 0.0cm 15.4cm 10cm, clip=true]{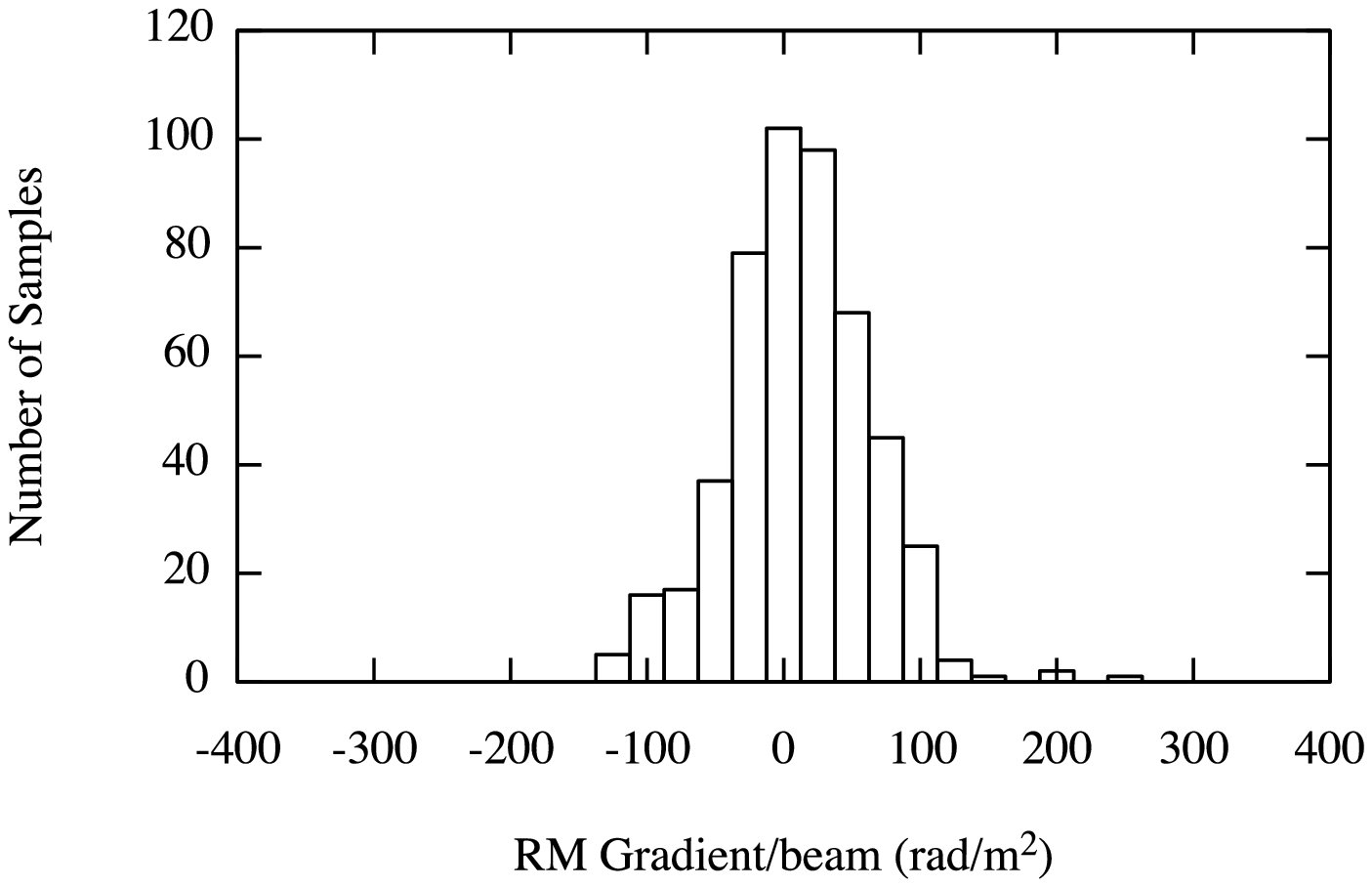}&
\includegraphics[width=5.5cm, trim=0cm 0.0cm 15.4cm 10cm, clip=true]{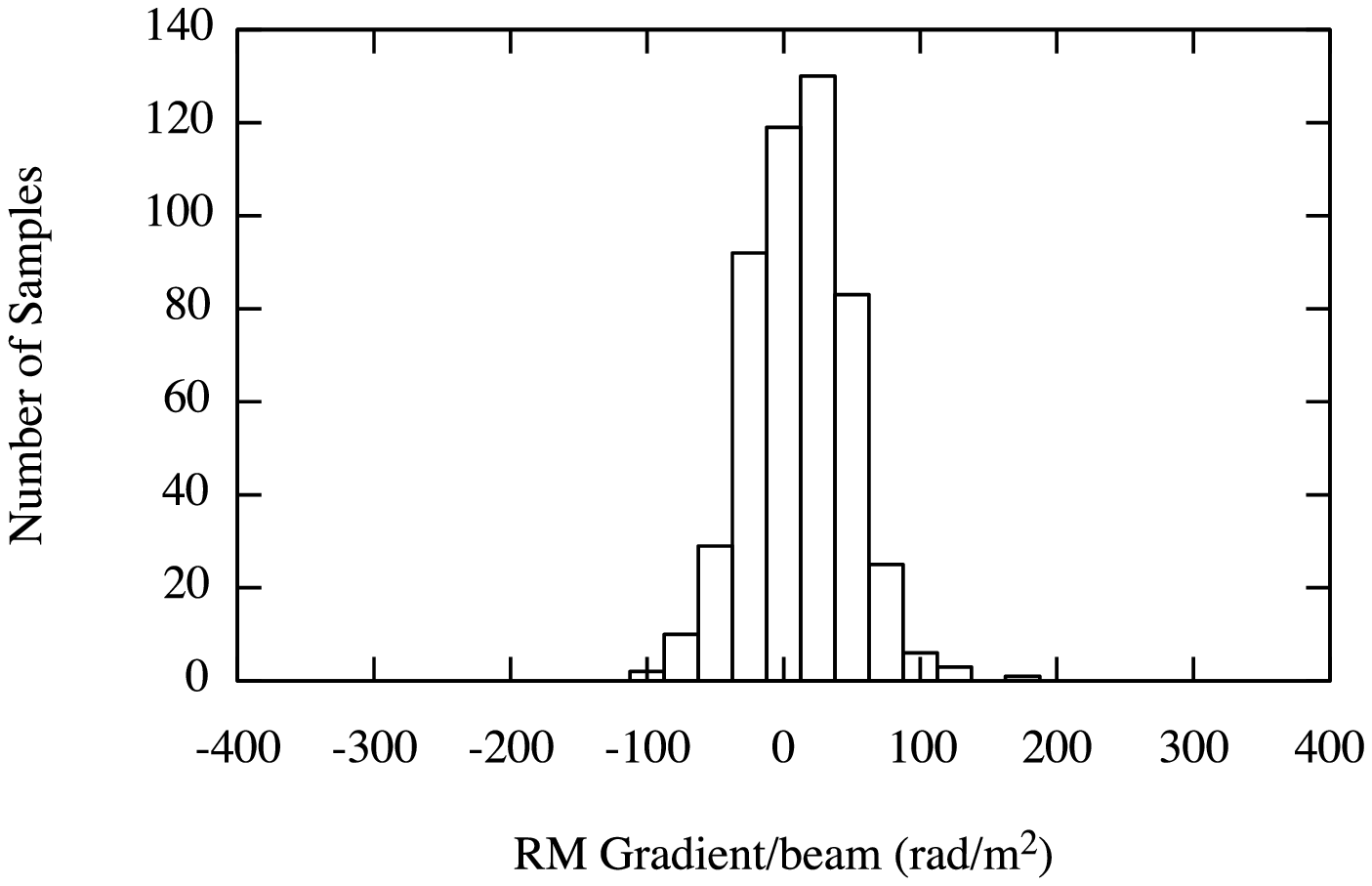}\\
(a) 1 beam &(b) 1.5 beams &(c) 2 beams\\

\includegraphics[width=5.5cm, trim=0cm 0.0cm 15.5cm 10cm, clip=true]{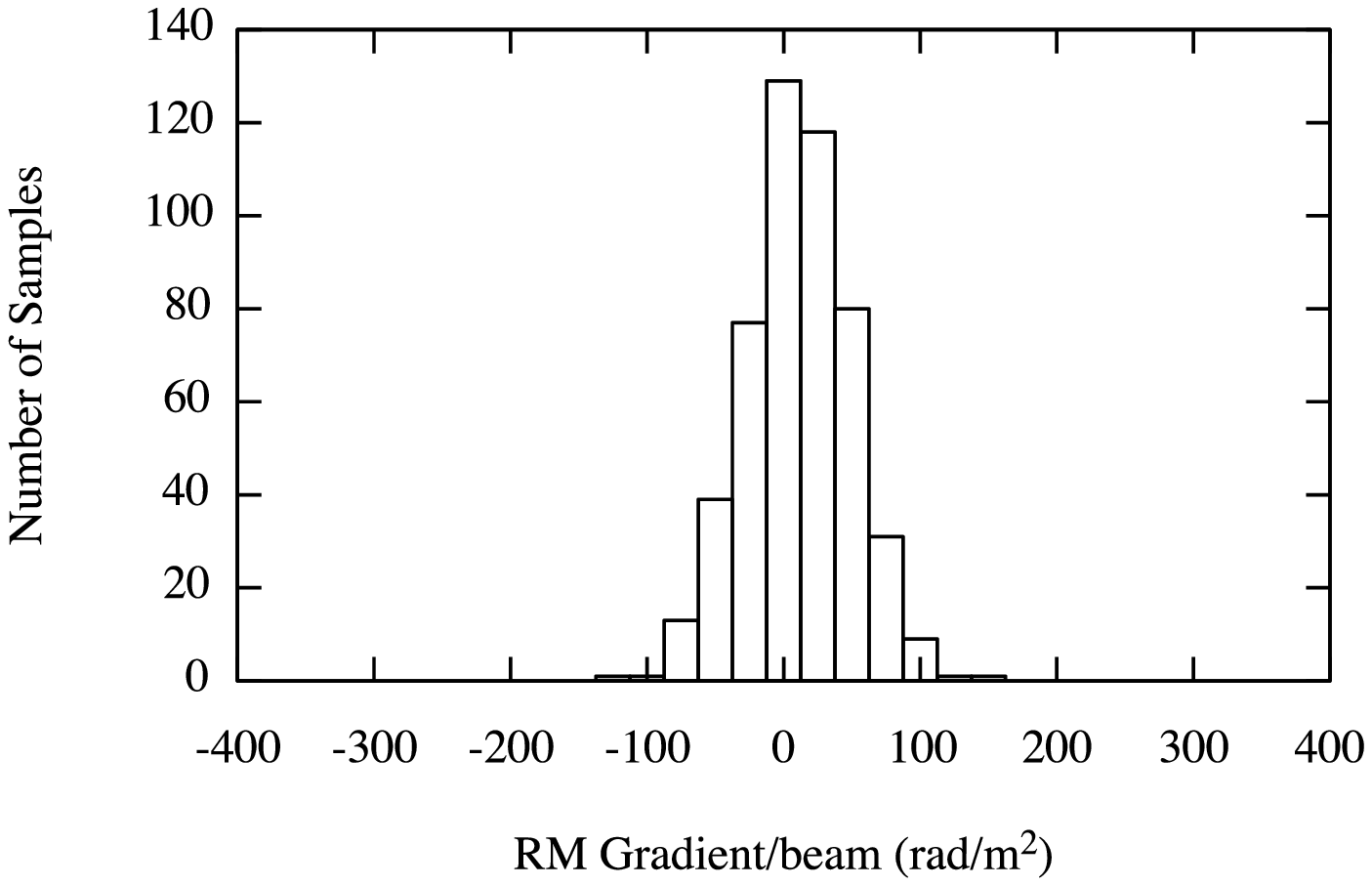}&
\includegraphics[width=5.5cm, trim=0cm 0.0cm 15.5cm 10cm, clip=true]{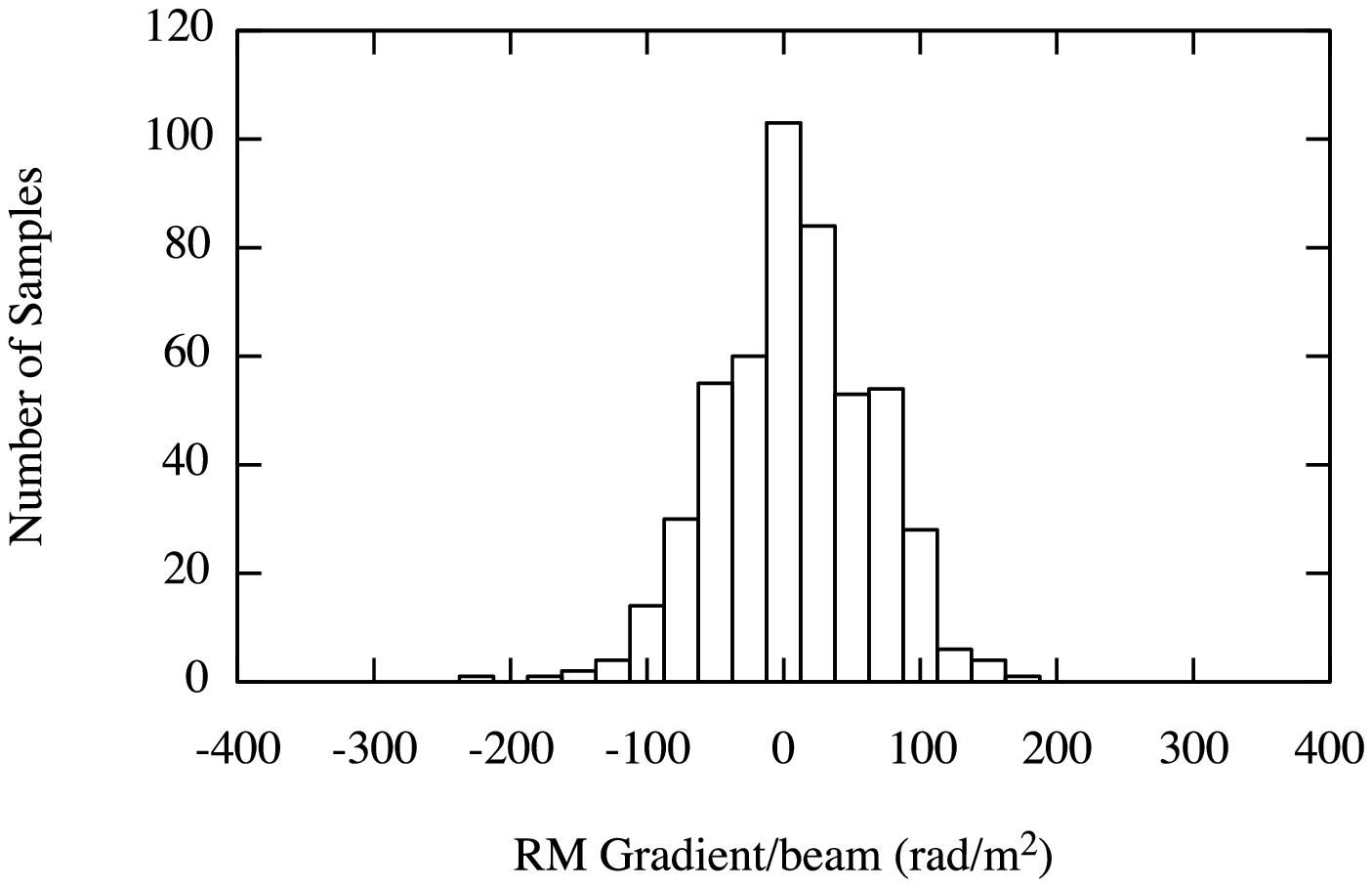}&
\includegraphics[width=5.5cm, trim=0cm 0.0cm 15.5cm 10cm, clip=true]{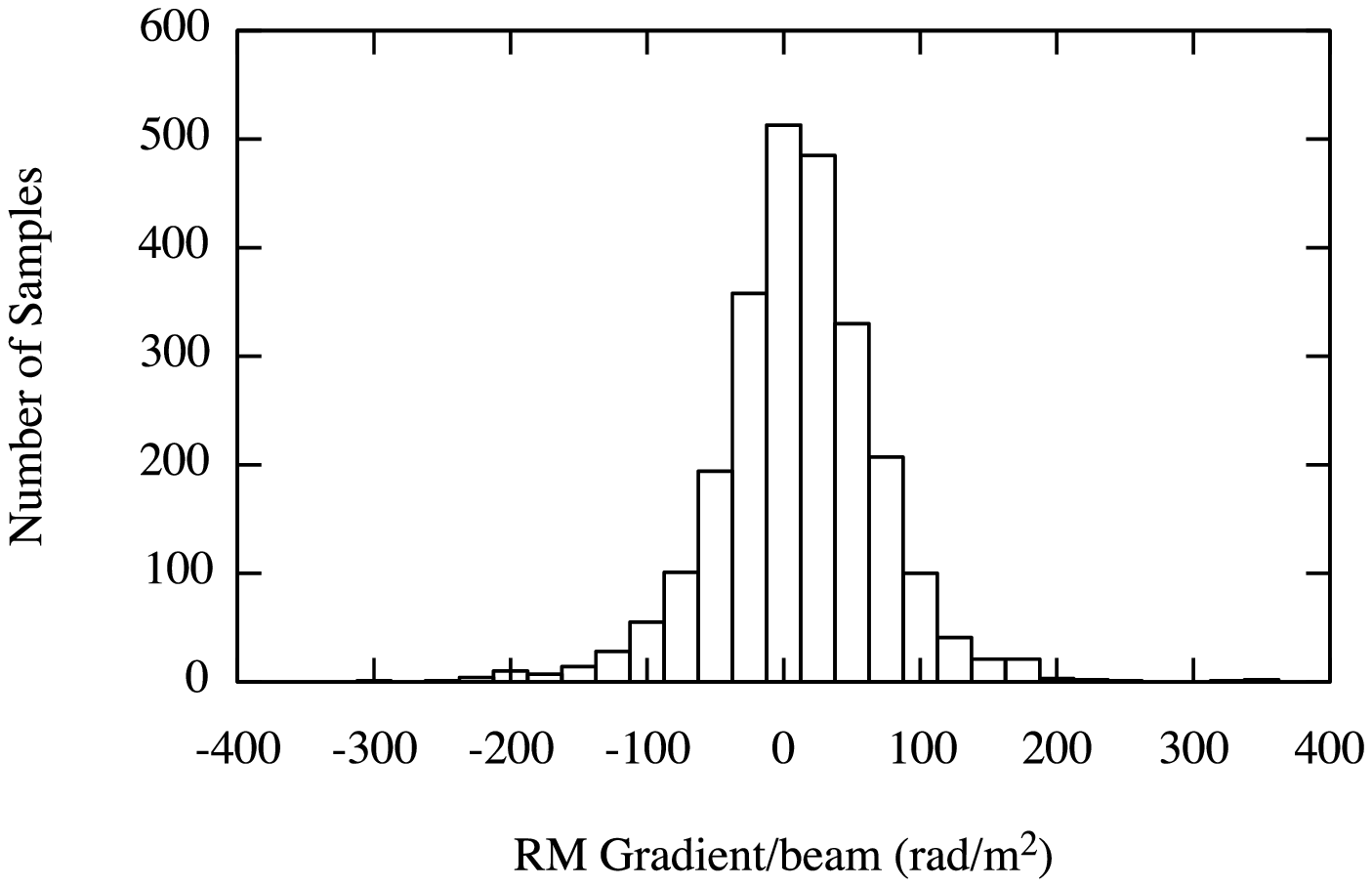}\\
(d) 2.5 beams &(e) 3 beams &(f) Total sample\\

  \end{tabular}
  \renewcommand{\thefigure}{A\arabic{figure}}%
\caption{Distributions of the 2500 simulated rotation measure gradients.}
\end{figure*}

\begin{figure*}
  \centering
  \begin{tabular}{ccc}
\includegraphics[width=5.5cm, trim=0cm 0.0cm 16cm 10cm, clip=true]{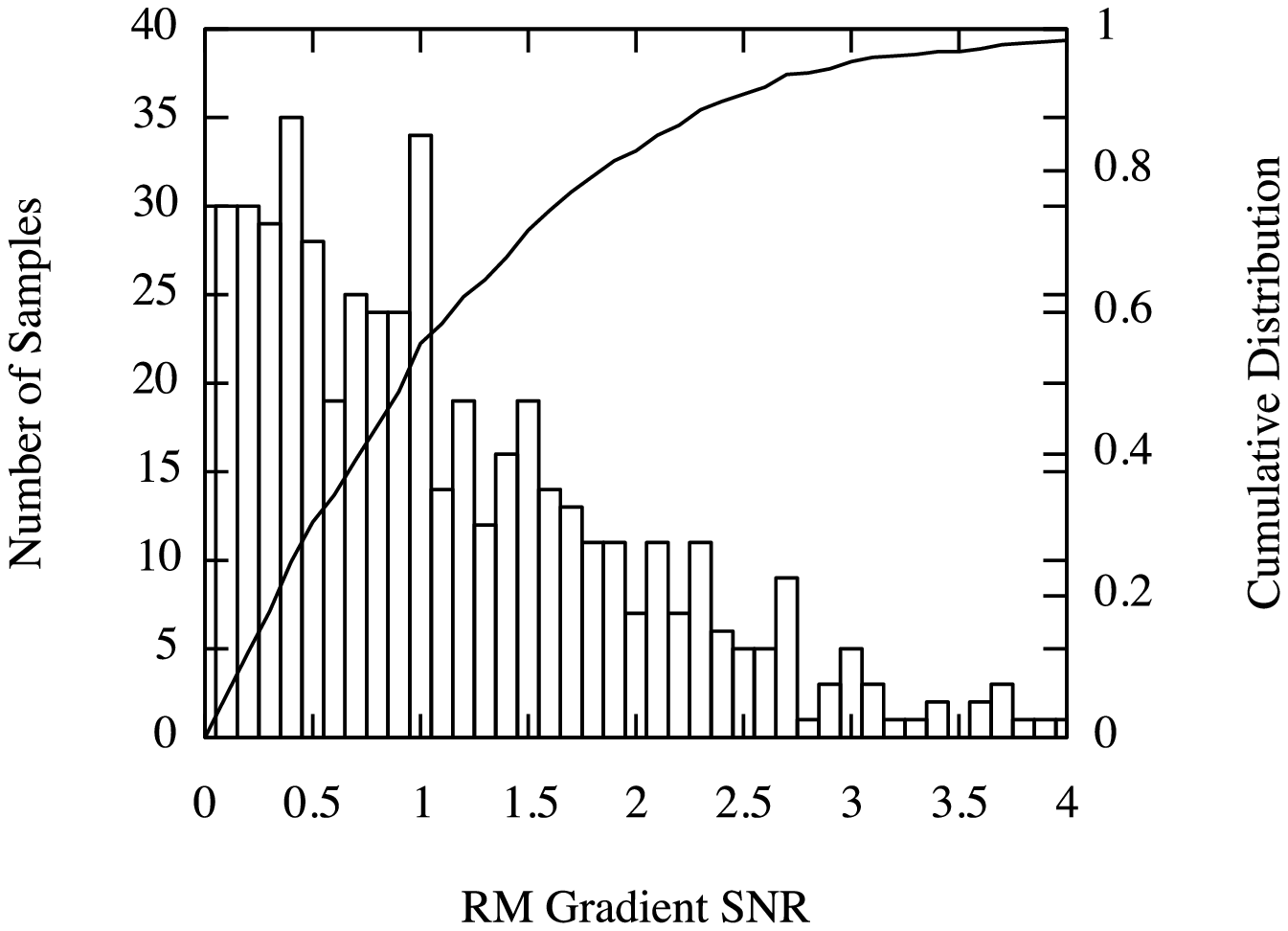}&
\includegraphics[width=5.5cm, trim=0cm 0.0cm 16cm 10cm, clip=true]{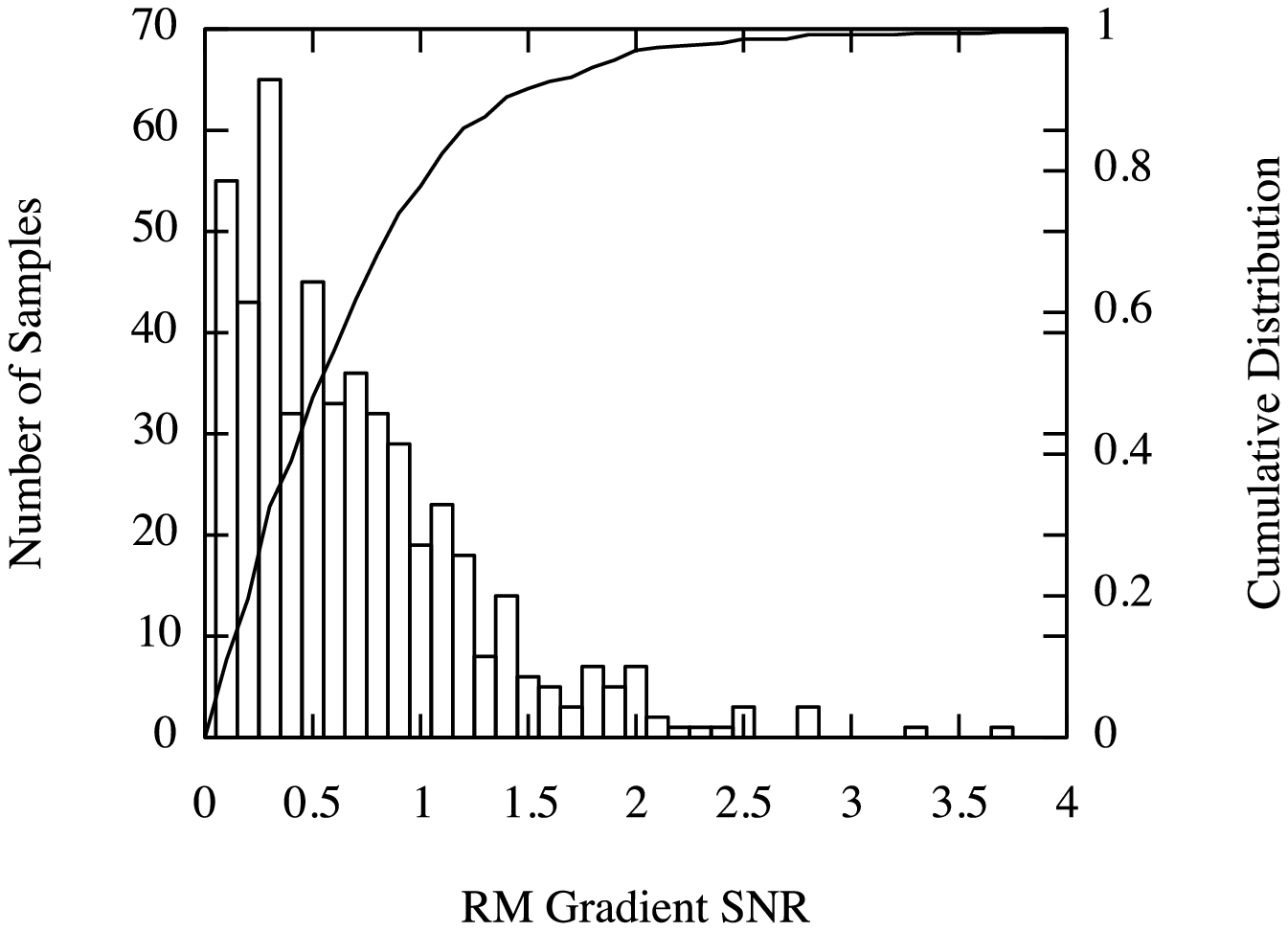}&
\includegraphics[width=5.5cm, trim=0cm 0.0cm 16cm 10cm, clip=true]{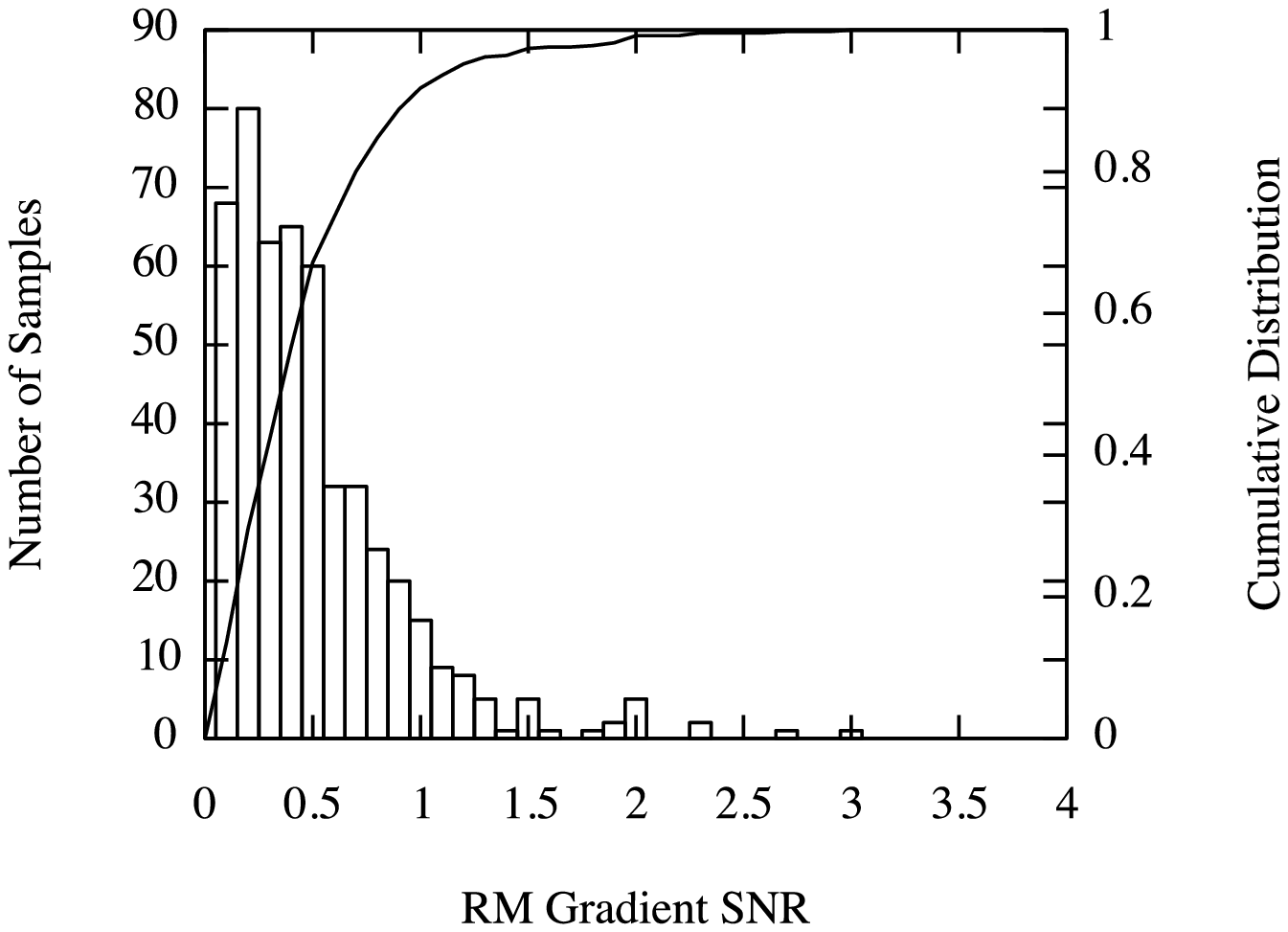}\\
(a) 1 beam &(b) 1.5 beams &(c) 2 beams\\

\includegraphics[width=5.5cm, trim=0cm 0.0cm 16cm 10cm, clip=true]{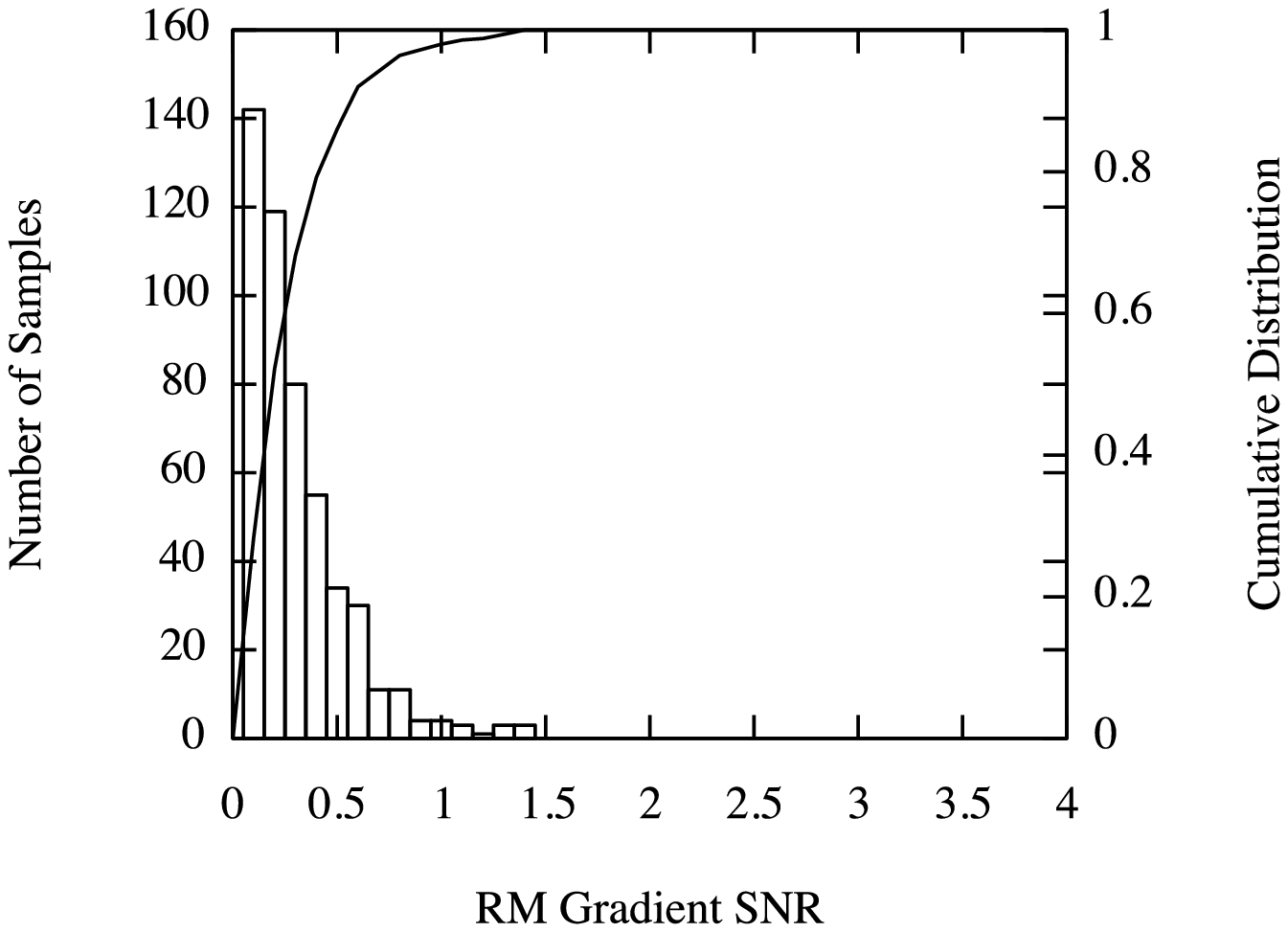}&
\includegraphics[width=5.5cm, trim=0cm 0.0cm 16cm 10cm, clip=true]{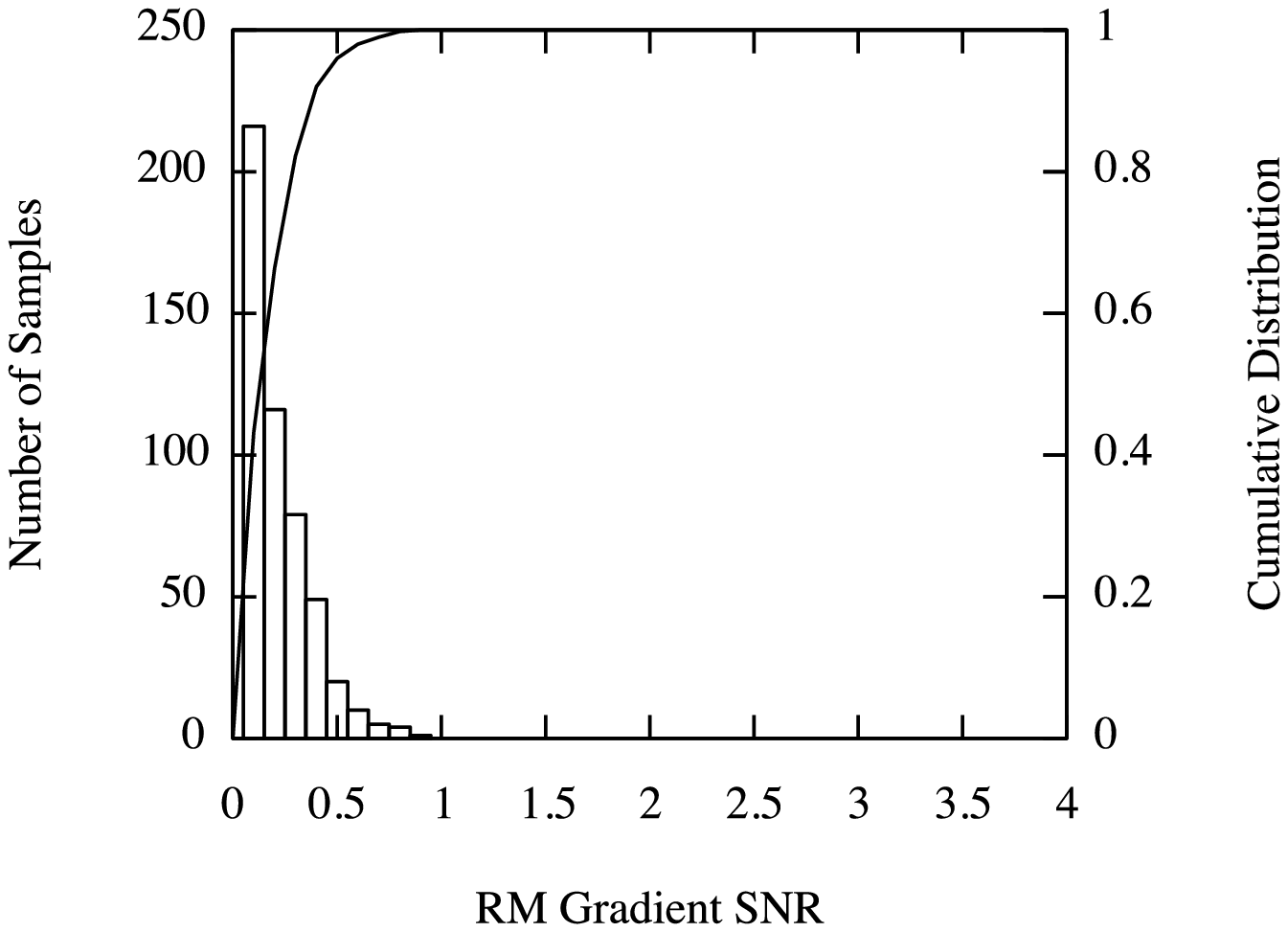}&
\includegraphics[width=5.5cm, trim=0cm 0.0cm 16cm 10cm, clip=true]{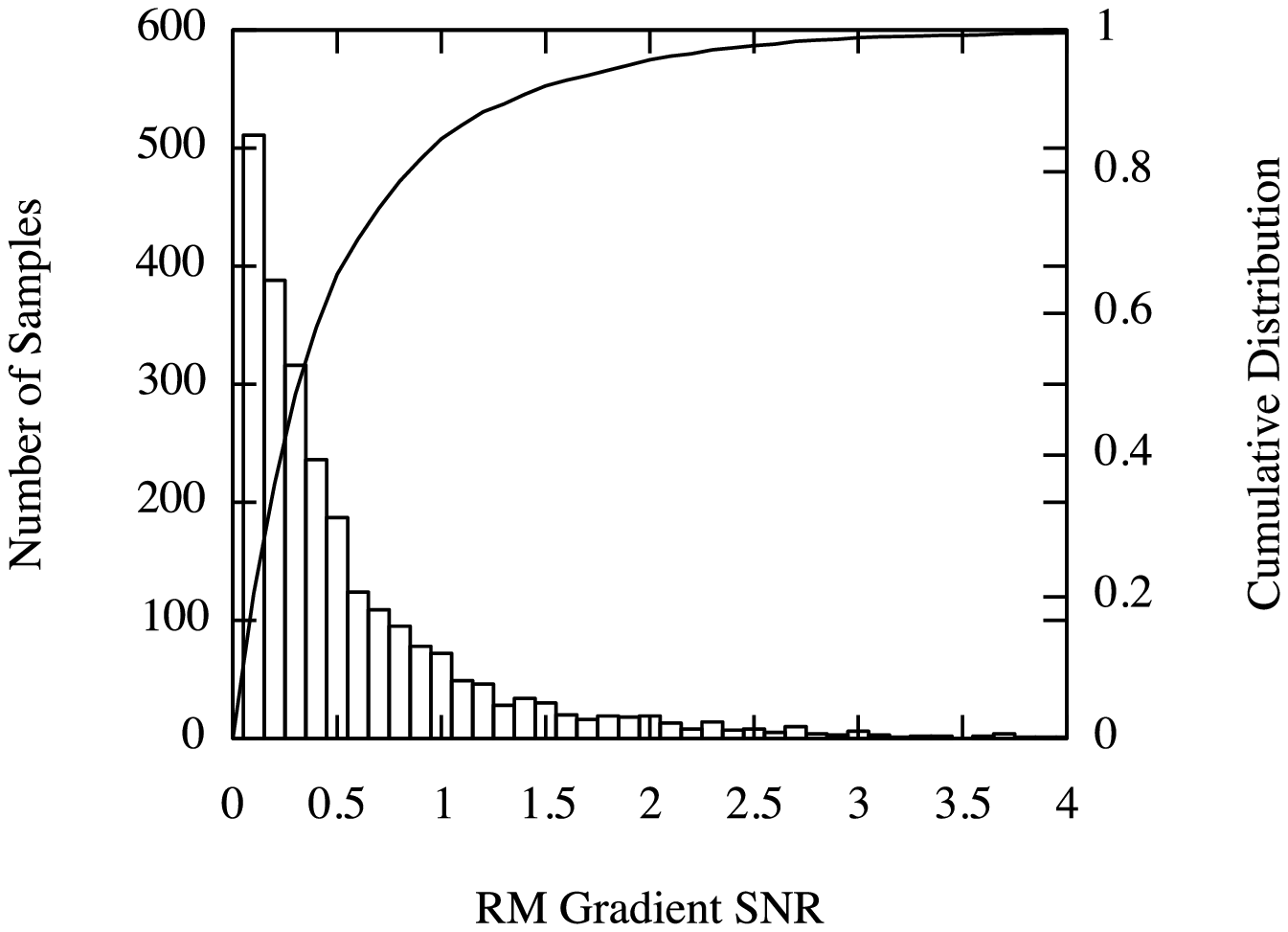}\\
(d) 2.5 beams &(e) 3 beams &(f) Total sample\\

  \end{tabular}
  \renewcommand{\thefigure}{A\arabic{figure}}%
\caption{Distributions of the 2500 simulated rotation measures gradients significance.}
\end{figure*}

\begin{figure}
  \centering
\includegraphics[width=8cm, trim=0.1cm 0.0cm 15cm 10cm, clip=true]{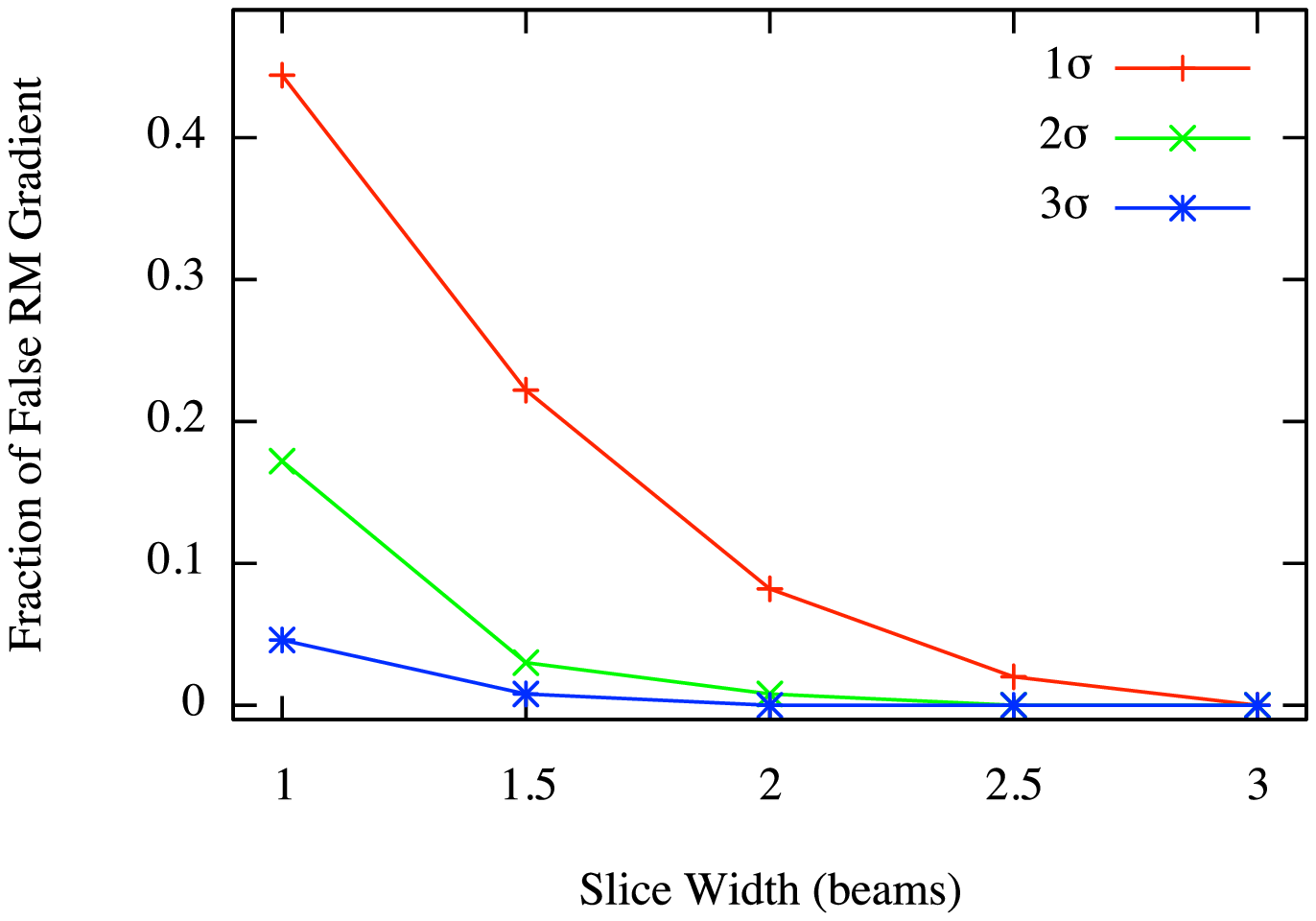}
\renewcommand{\thefigure}{A\arabic{figure}}%
\caption{Fraction of rotation measures along the slice in our simulation that exceeded 1 (``+'' symbol, red in the online version), 2 (``$\times$'' symbol, green in the online version) and 3 (``\varhexstar'' symbol, blue in the online version) standard deviations compared with size of the slice.}
\end{figure}

\section*{Acknowledgments}

This research has made use of data taken by the Very Long Baseline Array (VLBA). The VLBA is operated by the National Radio Astronomy Observatory (NRAO), and the NRAO is a facility of the National Science Foundation operated under cooperative agreement by Associated Universities, Inc. 
The author thanks M. Inoue, K. Asada, M. Nakamura and the anonymous referee for very useful comments that substantially improved this manuscript.

\label{lastpage}

\end{document}